%% file: paper.tex
\documentclass[submission,Phys]{SciPost}

\input{incl_settings.tex}   

\input{incl_shortcuts.tex}          
\graphicspath{{./figs/}}                

\begin{document}

\begin{center}{\Large \textbf{
Normalizing Flows for High-Dimensional Detector Simulations
}}\end{center}

\begin{center}
  Florian Ernst\textsuperscript{1,2},
  Luigi Favaro\textsuperscript{1,3},
  Claudius Krause\textsuperscript{1,4},
  Tilman Plehn\textsuperscript{1}, and 
  David Shih\textsuperscript{5}
  
\end{center}

\begin{center}
  {\bf 1} Institut f\"ur Theoretische Physik, Universit\"at Heidelberg, Germany \\
  {\bf 2} Experimental Physics Department, CERN, Geneva, Switzerland \\
  {\bf 3} CP3, Universit\'e catholique de Louvain, Louvain-la-Neuve, Belgium\\
  {\bf 4} Institut für Hochenergiephysik (HEPHY), Österreichische Akademie der Wissenschaften (ÖAW), Vienna, Austria\\
  {\bf 5} NHETC, Department of Physics \& Astronomy, Rutgers University, Piscataway, NJ USA 
\end{center}

\begin{center}
\today
\end{center}

\section*{Abstract}
  {Whenever invertible generative networks are needed for LHC physics,
    normalizing flows show excellent performance. In this work, 
    we investigate their performance for fast calorimeter shower 
    simulations with increasing phase space dimension. We use fast and expressive
    coupling spline transformations applied to the CaloChallenge
    datasets. In addition to the base flow
    architecture we also employ a VAE to compress the dimensionality
    and train a generative network in the latent space.
    We evaluate our networks on several metrics, including high-level features,
    classifiers, and generation timing.
    Our findings demonstrate that invertible neural networks have competitive performance when compared to autoregressive flows, while
    being substantially faster during generation. }

\vspace{-2pt}
\noindent\rule{\textwidth}{1pt}
\vspace{-20pt}
\tableofcontents\thispagestyle{fancy}
\noindent\rule{\textwidth}{1pt}
\vspace{0pt}

\clearpage
\section{Introduction}
\label{sec:intro}

Simulations are a defining aspect of LHC physics, bridging experiment
and fundamental theory and allowing for a proper interpretation of LHC
measurements~\cite{Campbell:2022qmc,Butter:2022rso}. The simulation
chain starts from the hard interaction at the scattering vertex, and
progresses through the radiation of soft particles, the decay of
heavy, unstable particles, hadronization of colored states, and
subsequent interaction of all particles in the event with the
detector. In the development of LHC as a precision-hadron collider,
the last step has become a major bottleneck in speed and precision, in
particular the reproduction of the detailed interactions of incident
and secondary particles within the calorimeters. Generating these
calorimeter showers with
\geant~\cite{Agostinelli:2002hh,1610988,ALLISON2016186}, based on
first principles, takes a substantial amount of the LHC computing
budget. Without significant progress, simulations will be the limiting
factor for all analyses at the high-luminosity upgrade of the LHC.

One development driving faster and more precise LHC simulations is the advent of deep generative networks. 
Fast detector simulations based on parametric models have been
extensively used in previous measurements from LHC experiments~\cite{ATLAS:2010arf, Rahmat:2012fs},
although their precision can be improved with machine learning
components~\cite{ATLAS:2021pzo}.
Such networks have shown great promise for LHC
physics in the past few years, providing fast and accurate surrogates
for simulations in high-dimensional phase spaces~\cite{Plehn:2022ftl}. They learn the
underlying probability distribution of events or calorimeter showers
from a reference dataset and then generate new samples based on this
learned distribution~\cite{Butter:2020qhk,Bieringer:2022cbs}. We have
seen successful applications to all steps in the simulation
chain~\cite{Butter:2022rso}, phase space
integration~\cite{Bendavid:2017zhk,Klimek:2018mza,Chen:2020nfb,Gao:2020vdv,Bothmann:2020ywa,Gao:2020zvv,Danziger:2021eeg,Heimel:2022wyj,Janssen:2023ahv,Bothmann:2023siu,Heimel:2023ngj};
parton
showers~\cite{deOliveira:2017pjk,Andreassen:2018apy,Bothmann:2018trh,Dohi:2020eda,Buhmann:2023pmh,Leigh:2023toe,Mikuni:2023dvk,Buhmann:2023zgc};
hadronization~\cite{Ilten:2022jfm,Ghosh:2022zdz,Chan:2023ume,Bierlich:2023zzd};
detector
simulations~\cite{Paganini:2017hrr,deOliveira:2017rwa,Paganini:2017dwg,Erdmann:2018kuh,Erdmann:2018jxd,Belayneh:2019vyx,Buhmann:2020pmy,ATL-SOFT-PUB-2020-006,Buhmann:2021lxj,Krause:2021ilc,ATLAS:2021pzo,Krause:2021wez,Buhmann:2021caf,Chen:2021gdz,Adelmann:2022ozp,Mikuni:2022xry,ATLAS:2022jhk,Krause:2022jna,Cresswell:2022tof,Diefenbacher:2023vsw,Hashemi:2023ruu,Xu:2023xdc,Diefenbacher:2023prl,Buhmann:2023bwk,Buckley:2023rez,Mikuni:2023tqg,Amram:2023onf,Diefenbacher:2023flw,FaucciGiannelli:2023fow,Pang:2023wfx};
and end-to-end event
generation~\cite{Otten:2019hhl,Hashemi:2019fkn,DiSipio:2019imz,Butter:2019cae,Alanazi:2020klf,Butter:2021csz,Butter:2023fov}.

For LHC physics, it is crucial that these networks are not used as
black boxes, but their performance can be investigated, understood,
and improved
systematically~\cite{Diefenbacher:2020rna,Butter:2021csz,Winterhalder:2021ave,Nachman:2023clf,Leigh:2023zle,Das:2023ktd}. This
is especially important when their conditional counterparts are used
for inference~\cite{Bieringer:2020tnw,Butter:2022vkj,Heimel:2023mvw},
probabilistic
unfolding~\cite{Datta:2018mwd,Bellagente:2019uyp,Andreassen:2019cjw,Bellagente:2020piv,Backes:2022vmn,Leigh:2022lpn,Raine:2023fko,Shmakov:2023kjj,Ackerschott:2023nax,Diefenbacher:2023wec},
or anomaly
detection~\cite{Nachman:2020lpy,Hallin:2021wme,Raine:2022hht,Hallin:2022eoq,Golling:2022nkl,Sengupta:2023xqy}.

In this paper, we will focus on the problem of building fast and accurate
surrogate models for calorimeter shower simulation, using the technology of normalizing flows. 
We have seen in a number of contexts that normalizing flows are a
promising technique for fast calorimeter simulation~\cite{Krause:2021ilc,Krause:2021wez,Krause:2022jna,Das:2023ktd}, but there are also major challenges with scaling them up to more granular (higher-dimensional) calorimeters~\cite{Diefenbacher:2023vsw,Buckley:2023rez,Cresswell:2022tof,Pang:2023wfx}.
These challenges are
especially interesting because recently, continuous-time generative models (diffusion models and continuous normalizing flows trained with flow-matching) have been tested on LHC
physics~\cite{Mikuni:2022xry,Butter:2023fov,Mikuni:2023tqg,Amram:2023onf,Heimel:2023mvw,Butter:2023ira,Leigh:2023toe,Mikuni:2023dvk,Buhmann:2023bwk,Buhmann:2023kdg,Leigh:2023zle,Buhmann:2023zgc,Birk:2023efj,Favaro:2024rle}
and show impressive performance which is not as limited by the dimensionality of the data. However, their gain in expressivity comes at
the expense of slower generation, leading to an interesting trade-off
between speed and quality of generated events or showers. 

Here, we will build on previous works \cite{Diefenbacher:2023vsw,Buckley:2023rez,Cresswell:2022tof,Pang:2023wfx} attempting to scale up normalizing flows to higher-granularity calorimeters. Focusing on the datasets~\cite{CaloChallenge_ds1,CaloChallenge_ds1_v3,CaloChallenge_ds2,CaloChallenge_ds3}
of the Fast Calorimeter Simulation Challenge~\cite{calochallenge,Krause:2024avx}, we will tackle this problem in two ways. 
\begin{itemize}
\item First, we will show how impressive gains in speed can be achieved by switching from the fully-autoregressive flows of ~\cite{Krause:2021ilc,Krause:2021wez,Krause:2022jna,Diefenbacher:2023vsw,Buckley:2023rez} to flows based on coupling layers\cite{2014arXiv1410.8516D,2016arXiv160508803D,glow} which are equally fast in the sampling and density estimation directions, while retaining or improving the network accuracy. Following the terminology of~\cite{inn,cinn}, we will refer to coupling-layer based flows as {\it invertible neural networks} (INNs) throughout this work. Using the INN framework, we are able to obtain state-of-the-art results on dataset~1 (pions and photons) and dataset~2 of the CaloChallenge.

\item Second, to reach the dimensionality of dataset~3, we will combine the INN framework with a VAE. Conceptually similar to other approaches, \cite{Cresswell:2022tof,esser2020disentangling,Toledo-Marin:2024gqh}, we will train the INN on the (much lower-dimensional) latent space of a VAE fit to the showers of dataset~3. Then sampling from the INN and passing this through the decoder of the VAE, we will obtain a surrogate model for dataset~3. We will see that the results here, while not state-of-the-art in terms of quality, are very fast to generate, so could fill out another point in the Pareto frontier of fast calorimeter shower simulation.

\end{itemize}

The paper starts by introducing the CaloChallenge datasets in
Sec.~\ref{sec:data}. In Sec.~\ref{sec:inn} we introduce our fast INN
version~\cite{inn,cinn} of a normalizing flow, as well as a VAE+INN
combination. In Sec.~\ref{sec:results} we discuss their performance on
the different dataset, with increasing phase space dimensionality and
including learned classifier weights. We conclude and provide timing
information in Sec.~\ref{sec:conclusion}. In the Appendices we provide
details on the different network architectures and hyperparameters and
compare the INN performance to CaloFlow.

\section{Datasets}
\label{sec:data}

\begin{table}[b!]
    \centering
    \begin{small} \begin{tabular}{l|c|ccccc}
      \toprule
      $E_\text{inc}$ & 256~MeV~...~131~GeV & 262~GeV   & 0.524~TeV  & 1.04~TeV  & 2.1~TeV   & 4.2~TeV  \\
      \midrule
      photons       & 10000 per energy & 10000  & 5000  & 3000  & 2000  & 1000 \\
      pions         & 10000 per energy & 9800   & 5000  & 3000  & 2000  & 1000 \\
      \bottomrule
    \end{tabular} \end{small}
    \caption{Sample sizes for different incident energies in
      dataset~1.}
    \label{tab:Einc.ds1}
\end{table}

As stated above, our reference datasets are the public datasets of the
Fast Calorimeter Simulation Challenge~\cite{calochallenge,Krause:2024avx} and represent
three increasing dimensionalities from the current LHC calorimeter 
granularity to the ultra high granularity of future calorimeters 
proposed for ILC~\cite{Bambade:2019fyw}, CLIC~\cite{Aicheler:2018arh}, FCC~\cite{FCC:2018evy} and beyond.
We use the public
datasets~\cite{CaloChallenge_ds1,CaloChallenge_ds1_v3,CaloChallenge_ds2,CaloChallenge_ds3}
of the Fast Calorimeter Simulation Challenge~\cite{calochallenge}.
They consist of showers simulated with \geant for different
incident particles.  The general geometry is the same across all
datasets: the detector volume is segmented into layers in the
direction of the incoming particle. Each layer is segmented
along polar coordinates in radial ($r$) and angular ($\alpha$) bins. A shower is given
as the incident energy of the incoming particle and the energy
depositions in each voxel.\medskip

Dataset~1 (DS1) provides calorimeter showers for central photons and
charged pions. They have been used in
\textsc{AtlFast3}~\cite{ATLAS:2021pzo}.  The voxelizations of the 5
photon layers and 7 pion layers in radial and angular  bins ($n_r\times
n_\alpha$) are
\begin{align}
  \text{photons} &\qquad
  8\times1, \; 16\times10, \; 19\times10, \; 5\times1, \; 5\times1 \notag \\
  \text{pions} &\qquad 
  8\times1, \; 10\times10, \; 10\times10, \; 5\times1, \; 15\times10, \; 16\times10, \; 10\times1
\label{eq:voxels_ds1}
\end{align}
This gives 368 voxels for photons and 533 voxels for pions. The incoming
particles are simulated for 15 different incident energies
$E_\text{inc} = 256~\text{MeV}~...~4.2~\text{TeV}$, increasing by 
factors of two, with the sample sizes given in Tab.~\ref{tab:Einc.ds1}.
The original ATLAS dataset does not require an energy threshold. The
effect of a threshold on the shower distributions at the detector cell
level requires further studies. We require $E_\text{min}=1$~MeV to all
generated voxels, motivated by the readout threshold of the
calorimeter cells and the fact that photon showers require a minimum
cell energy of 10~MeV to cluster and pion showers start clustering at
300~MeV~\cite{MicheleFaucciGiannelli}. We note that the usually called $E_{\text{inc}}$ is in reality the momentum of the incoming particle. This has implications for pions which have to be further studied for a future deployment.\medskip

Datasets~2 and~3 (DS2/3) are not modeled after existing
detectors. They assume 45~layers of active silicon detector (thickness
0.3~mm), alternating with inactive tungsten absorber layers (thickness
1.4~mm) at $\eta=0$.  Each dataset contains 100,000 \geant electron showers
with log-uniform $E_\text{inc} = 1~...~1000$~GeV.  The only
difference between the two datasets is the voxelization. In dataset~2,
each layer is divided into $16 \times 9$ angular and radial voxels,
defining 6480~voxels in total. Dataset~3 uses $50 \times 18$ voxels per
layer or 40,500 voxels in total. The minimal recorded energy per voxel
for these two datasets is $15.15$~keV.

\section{CaloINN}
\label{sec:inn}

We study two different network architectures. First, we benchmark a
standard INN and demonstrate its precision and generation speed
especially for low-dimensional phase space. Second, we embed this INN
in a VAE, with the goal of describing datasets~2 and~3 with the same
physics content, but a much larger phase space dimensionality.

\subsection{INN}

Normalizing flows describe bijective mappings between a (Gaussian)
latent space $r$ and the physical phase space $x$,
\begin{align}
\pl(r)
\quad
\stackrel[\leftarrow \; \overline{G}_\theta(x)]{G_\theta(r) \rightarrow}{\xleftrightarrow{\hspace*{1.5cm}}}
\quad
\pmd(x) \sim \pd(x) \; .
\label{eq:inn_mapping}
\end{align}
$\overline{G}_\theta(x)$ denotes the inverse transformation to
$G_\theta(r)$. The INN variant~\cite{inn,cinn} of normalizing flows is
completely symmetric in the two directions. After training the
network,\\
$\pd(x) \sim \pmd(x)$, we use the INN to sample $\pmd(x)$
from $\pl(r)$~\cite{Plehn:2022ftl}.

The building block of our INN architecture is the coupling layer \cite{2014arXiv1410.8516D,2016arXiv160508803D,glow}. It allows for a Jacobian calculable in a single network evaluation for both $\overline{G}_\theta(x)$ and $G_\theta(r)$.
Therefore, we train the INN with a likelihood loss
\begin{align}
  \loss_\text{INN}
  = - \XLangle \log \pmd(x) \XRangle_{\pd} 
  =- \XXLangle \log \pl\big(\overline{G}_\theta(x)\big) + \log \left| \frac{\partial \overline{G}_\theta(x)}{\partial x}\right|
  \XXRangle_{\pd} \; .
  \label{eq:MLE}
\end{align}
The first term ensures that the latent representation remains
Gaussian, while the second term constructs the correct transformation
to the phase space distribution. Given the structure of
$\overline{G}_\theta(x)$ and the latent distribution $\pl$, both terms
can be computed efficiently.

Figure~\ref{fig:schematic} (left) shows a schematic representation of the CaloINN architecture.
In the coupling block we split the input vector, consisting of the normalized voxels $x$ and the energy variables $u$, in two equally-sized vectors.
The first half of the vector is not transformed and used, together with the logarithm
of the incident energy $E_\text{inc}$, to predict the parameters of the
transformation applied to the second half.

A standard affine transformation uses a
scale and shift parameter for each voxel. Instead, we define a spline parametrized by a neural network.
We employ rational quadratic splines (RQS) and cubic splines. 
The splines are defined piecewise in a box of size $[-B, B]$.
Given the total number of bins $K$, the spline is parametrized by the locations of each knot and their first derivatives. 
In one bin $k$, the two transformations have the form
\begin{equation}
f_\text{RQS}(x)_k = \frac{\alpha_{0,k} + \alpha_{1,k} x + \alpha_{2,k} x^2}
            {\beta_{0,k} + \beta_{1,k} x + \beta_{2,k} x^2}
\qquad \text{and} \qquad 
f_\text{cubic}(x)_k = \gamma_{0,k} + \gamma_{1,k} x + \gamma_{2,k} x^2 + \gamma_{3,k} x^3 \, ,
\end{equation}
respectively. The parameters $(\alpha, \beta)_k$ and $(\gamma)_k$
can be expressed as a function of the bin height, width, and first derivative in a stable numerical form.
The complete parametric expression for the RQS can be found in~\cite{durkan2019neural}, while the implementation of the cubic spline follows~\cite{durkan2019cubic}.
The total number of parameters predicted by the neural network,
after accounting for the continuity constraints, are $3K -1$ for each transformed variable.
The large scale architecture stacks several coupling blocks each one followed by a permutation of the input and an ActNorm~\cite{glow} layer
for normalization purposes.
The INN is implemented using the \textsc{FrEIA}\footnote[4]{We provide the code in a Github repository at \href{https://github.com/heidelberg-hepml/CaloINN}{https://github.com/heidelberg-hepml/CaloINN}} package~\cite{freia}.
For datasets~1, we employ a
rational quadratic spline, while for dataset 2
we find cubic splines to give more stable results.  A discussion on the ablation studies is provided in App.~\ref{app:hyperparameters} together with all the INN hyperparameters.

As a noteworthy preprocessing we normalize each shower to the layer
energy. The energy information is encoded as
\begin{align}
  u_0 = \frac{\sum_i E_i}{E_\text{inc}}
  \qqquad \text{and} \qqquad 
  u_i = \frac{E_i}{\sum_{j\ge i} E_j} \; ,
\label{eq:enc_energy}
\end{align}
in terms of the energy depositions per layer $E_i$. The $u_i$ are
appended to the list of voxels for each shower. We do not explore a 
separate training for the energy and the voxel dimensions which would
simplify the learning process of the energy dimensions. We train the INN on
the full data, conditioned on the logarithm of the incident energies.
Unlike, for instance,
CaloFlow~\cite{Krause:2021wez,Krause:2022jna,Buckley:2023rez} we train
a single network without any distillation. We provide the details
of the preprocessing in App.~\ref{app:hyperparameters}.

\begin{figure}
    \centering
    \scalebox{.5}{\input{figs/networks/inn}} \hspace{2cm}
    \scalebox{.5}{\input{figs/networks/vae-inn}}
\caption{Schematic representation of the CaloINN (left) and the CaloVAE+INN (right) architectures.}
\label{fig:schematic}
\end{figure}
%

\subsection{VAE+INN}

The problem with the INN is the scaling towards dataset~3 with its
high-dimensional phase space of 40k voxels. The INN scales at least linearly in time and memory with the input dimension since each voxel is processed by a spline that has to be parameterized independently. In practice, the scaling is usually worse than linear, as the number of parameters, needed to parameterize each spline, tends to grow with the number of voxels as well. To solve this scaling
problem we introduce an additional VAE to reduce the dimensionality of
the INN mapping. Differently from ~\cite{Cresswell:2022tof}, we do not estimate the dimensionality of the manifold but rather optimize the reconstruction of the VAE while keeping a low-dimensional latent space. The VAE consists of a
preprocessing block, an encoder-decoder combination, and a
postprocessing block.  Both, the decoder and the encoder are
conditioned on the incident energies and additional energy
variables. Therefore, we compress normalized showers in the latent
space and jointly learn the energy and the latent variables with the
INN. During generation, the INN samples into the latent space of the
VAE, and the VAE decoder translates this information to the shower
phase space. We set the latent space to 50 for dataset~1 and dataset~2, and to 300 for dataset~3. Other specifics of the network are different in the three datasets and are provided in App.~\ref{app:hyperparameters}.

The goal of our $\beta$-VAE \cite{higgins2017betavae, Burgess:2018hqi} is to learn to reconstruct the
input data. We assume a Gaussian distribution for the encoder network
$E(z|x)$.  The VAE loss for the compression is
\begin{align}
\loss = \loss_\text{BCE} + \beta \kl [ E(z|x), \pl(z)] \; ,
\label{eq:vae_loss}
\end{align}
with the usual binary cross entropy loss and the Gaussian prior.  For
a Gaussian encoder the KL-divergence can be computed analytically, and
the coupling strength is $\beta=10^{-9}$. We select this small value as the KL part is only a regularization in our setup. We do not need a Gaussian latent space since a more expressive mapping is learned by the INN. However, we need very accurate decoding abilities from the latent space. The only requirement, that is ensured by the small KL term, is a compact well-behaved space which can be learned by a generative model. For numerical stability we split it into an upscaling factor for the BCE part and a downscaling factor for the KL part.

For the decoder we use a Bernoulli likelihood, because it outperforms
other models. For example the Gaussian and the continuous Bernoulli \cite{loaizaganem2019continuous} approach. The Gaussian decoder does not
model the shower geometry well, and it under-populates the low-energy
regions. The continuous Bernoulli distribution leads to instabilities,
as the average energy deposition in the normalized space is close to
zero.  We use a Bernoulli decoder,
\begin{align}
D(x|\lambda(z)) = \lambda(z)^x (1-\lambda(z))^{1-x}  \; ,
\end{align}
defining the combined VAE loss
\begin{align}
\loss_\text{VAE} 
= \XXLangle \XLangle x \log \lambda + (1-x) \log\left(1-\lambda \right)  \XRangle_{z \sim E(z|x)}
+ \beta \left[1+\log \sigma_E^2
-\mu_E^2 -\sigma_E^2 \right] \XXRangle_{x \sim \pd} \; .
\label{eq:bvae_loss}
\end{align}
Because the Bernoulli distribution gives a binary probability we use
its continuous mean $\lambda$ as the prediction for the individual
voxels.

The remaining differences between the unit-Gauss prior in the latent
space and the encoder are mapped by the INN.  Applying a 2-step
training, we first train the VAE and then train the INN given the
learned latent space. This means we pass the
encoder means and the standard deviations, as well as the energy
variables to the INN. The INN is trained as described above, mapping
the latent representation of the VAE to a standard Gaussian.  As for
the full INN, the energy information is encoded following
Eq~\eqref{eq:enc_energy} and learned by the latent flow. Both encoder
and decoder of the VAE are conditioned to these variables.

For the larger datasets~2 and 3, we employ a mixture of a
convolutional and a fully connected VAE.  Our assumption is that the
calorimeter layers do not require a full correlation, but that only neighboring layers are strongly correlated. This assumption is simply implementing locality, which should be given due to the causal propagation of the shower through the calorimeter layers and the regular structure of electromagnetic showers. Therefore, we can simplify the
structure by compressing consecutive layers jointly in a first-step
compression. We use an architecture with fully connected
sub-blocks, resembling a kernel architecture with a kernel size $k$
(number of jointly encoded calorimeter layers) and a stride $s$
(distance between two neighboring kernel blocks). After this first
compression we concatenate these latent sub-spaces and compress them
a second time into our final latent space. For the decoding we reverse
this two-step structure. The overlapping regions of the fully
connected kernel blocks are summed over. It should be noted, that the large scale correlations are not completely ignored in this approach. They can still be learned as correlations between the kernel blocks in the second stage. However, it is harder to model them as the information content is already compressed by the first dimensionality reduction.

\section{Results}
\label{sec:results}

The main physics reason for specific shower features is the incident
energy.  Low-energy showers will interact with only a few layers of
the calorimeter and quickly widen, leading to a broad center-of-energy
distribution in earlier calorimeter layers and a high sparsity in the
given voxelization.  High-energy showers penetrate the calorimeter
more deeply. They will be collimated in the initial layers and  have low
sparsity since each shower is likely to deposit energy in each voxel.

To see if the ML-learned showers reflect these physics properties, we
look at physics-motivated and high-level features. Given a shower with energy depositions ${\mathcal I}$, we look at the center of
energy and its width for each layer, 
\begin{align}
    \langle \zeta \rangle = \frac{\mathbf{\zeta} \cdot \mathbf{\mathcal I}}{\sum_i {\mathcal I}_i}
    \qquad \text{and} \qquad 
    \sigma_{\langle \zeta \rangle} = \sqrt{\frac{\mathbf{\zeta^2} \cdot \mathcal{I}}{\sum_i {\mathcal I}_i} 
                    - \langle \zeta \rangle^2} 
    \qquad \text{for} \qquad 
    \zeta = \eta, \phi \; ;
\end{align}
where $\sum_i$ runs over the voxels in one layer. We also look at
the energy deposition in each layer; the layer sparsity; and for dataset~1, 
the ratio $E_\text{tot}/E_\text{inc}$ for each discrete incident energy.\medskip

To analyze the quality of our generative networks in more detail and
to identify failure modes, we train a classifier $D(x)$ on the voxels,
to distinguish \geant showers from generated
showers~\cite{Krause:2021ilc, Krause:2021wez,Das:2023ktd}.  By the
Neyman-Pearson lemma the trained classifier approximates the
likelihood-ratio.  This means we can compute the correction
weight~\cite{Das:2023ktd} and use the weight distributions as an
evaluation metric
\begin{align}
w(x) = \frac{D(x)}{1-D(x)} \approx \frac{p_\text{data}}{p_\text{model}}(x) \; .
\label{eq:def_weights}
\end{align}
For these weights it is crucial that we evaluate them on the training
and on the generated datasets combined, because typical failure modes
correspond to tails for one of the two datasets~\cite{Das:2023ktd}. In
addition, we always check if showers with especially small or large
weights cluster in phase space, allowing us to identify failure modes
of the respective generative network.
We also report the Area-Under-the-Curve (AUC) score, which is calculated after training ten classifiers from different initializations and averaging the obtained AUC scores.

\subsection{Dataset 1 photons}

\begin{figure}[b!]
    \includegraphics[width=0.33\textwidth, page=1]{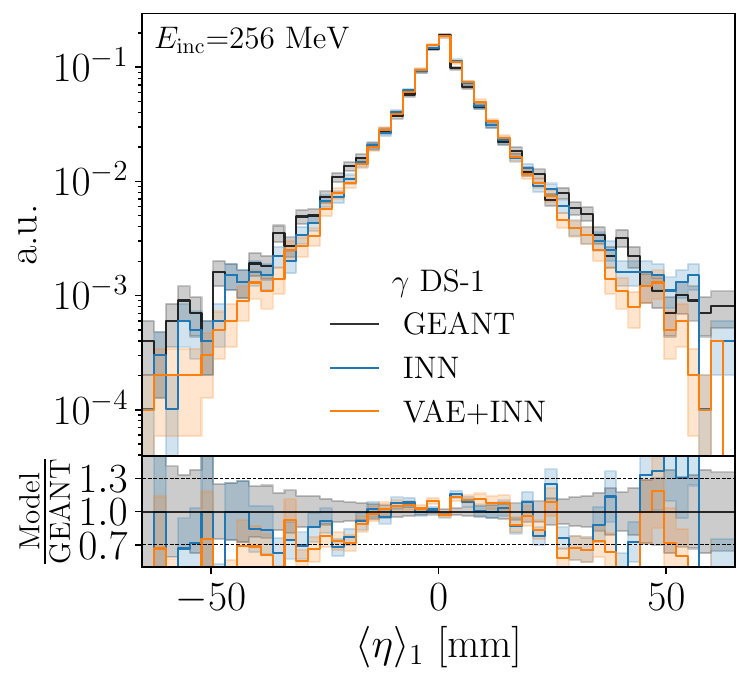}
    \includegraphics[width=0.33\textwidth, page=2]{figs/ds1-photons/E_8192/ECEta_layer_dataset_1-photons.pdf}
    \includegraphics[width=0.33\textwidth, page=2]{figs/ds1-photons/E_262144/ECEta_layer_dataset_1-photons.pdf} \\
    \includegraphics[width=0.33\textwidth, page=1]{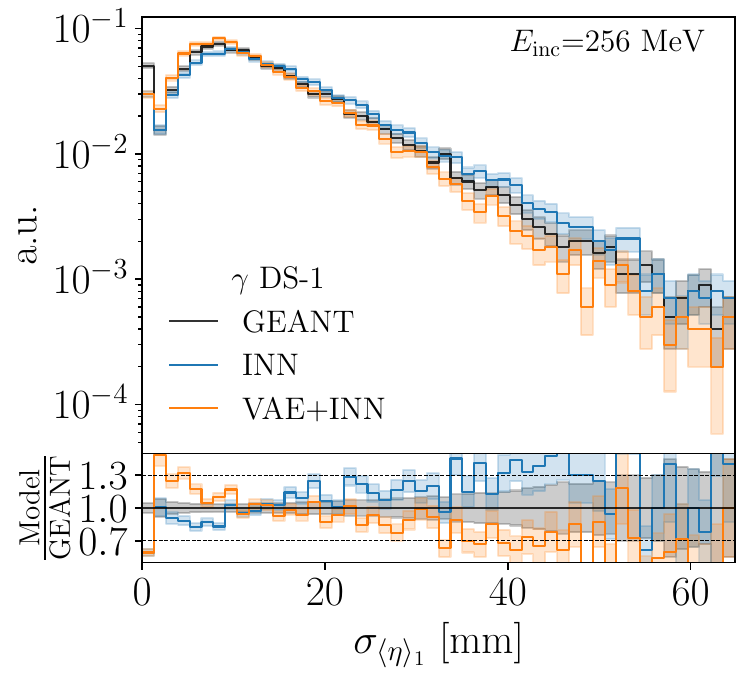}
    \includegraphics[width=0.33\textwidth, page=2]{figs/ds1-photons/E_8192/WidthEta_layer_dataset_1-photons.pdf}
    \includegraphics[width=0.33\textwidth, page=2]{figs/ds1-photons/E_262144/WidthEta_layer_dataset_1-photons.pdf}\\
    \includegraphics[width=0.33\textwidth, page=2]{figs/ds1-photons/E_256/E_layer_dataset_1-photons.pdf}
    \includegraphics[width=0.33\textwidth, page=2]{figs/ds1-photons/E_8192/E_layer_dataset_1-photons.pdf}
    \includegraphics[width=0.33\textwidth, page=2]{figs/ds1-photons/E_262144/E_layer_dataset_1-photons.pdf} \\
    \includegraphics[width=0.33\textwidth]{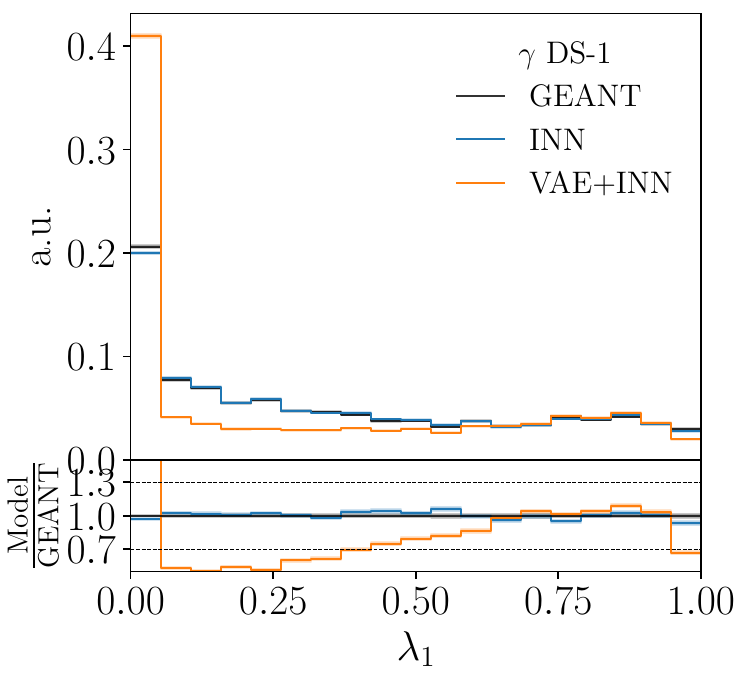}
    \includegraphics[width=0.33\textwidth]{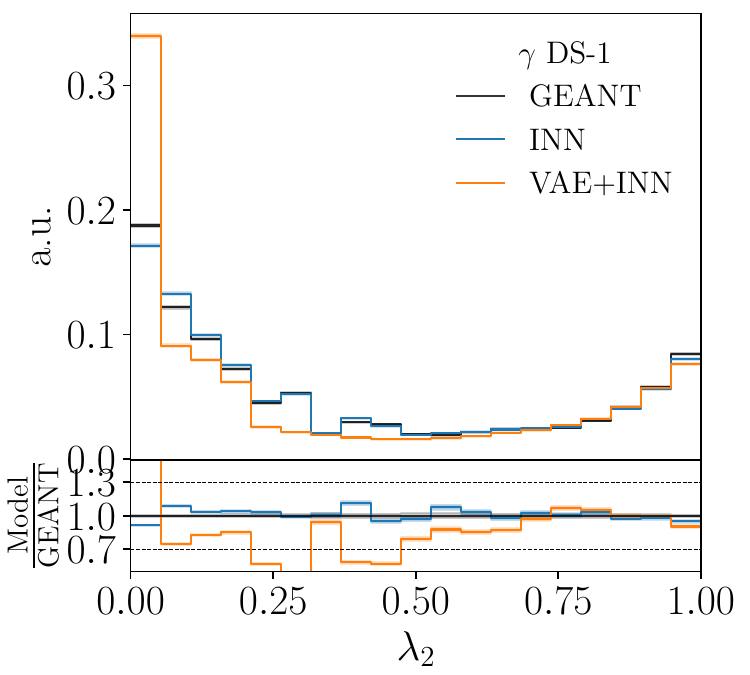}
    \includegraphics[width=0.33\textwidth]{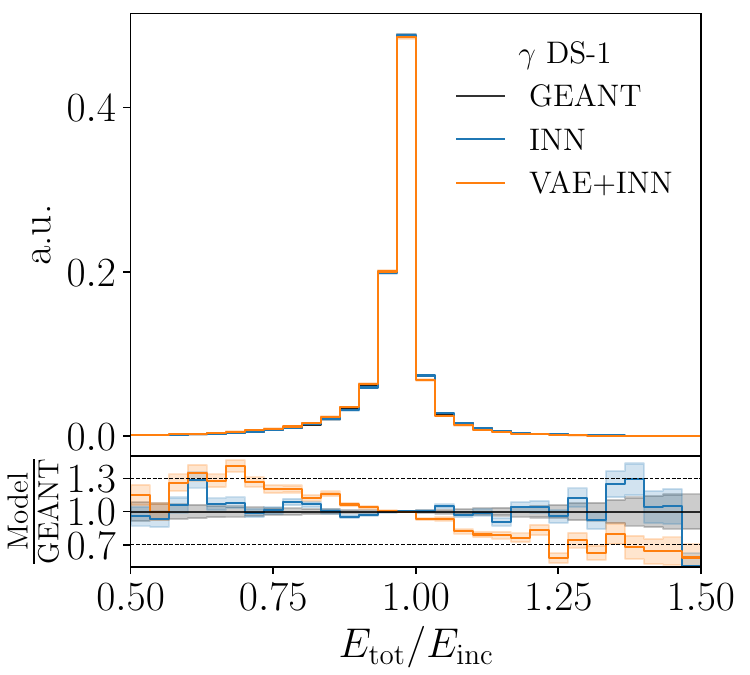}
    \caption{Set of high-level features for $\gamma$ showers in
      dataset~1, compared between \geant, INN, and VAE+INN. 
      We show the energy deposition, the center of energy, and the 
      width of the center of energy in layer-1 for the incident energies 256 MeV, 8.2 GeV, and 262.1 GeV. The last row contains the inclusive sparsity in layer-1 and layer-2, and the inclusive
      energy ratio $E_\text{tot}/E_\text{inc}$.}
    \label{fig:hlf_ds1_photons}
\end{figure}

\begin{figure}[t!]
    \centering
    \includegraphics[width=\textwidth]{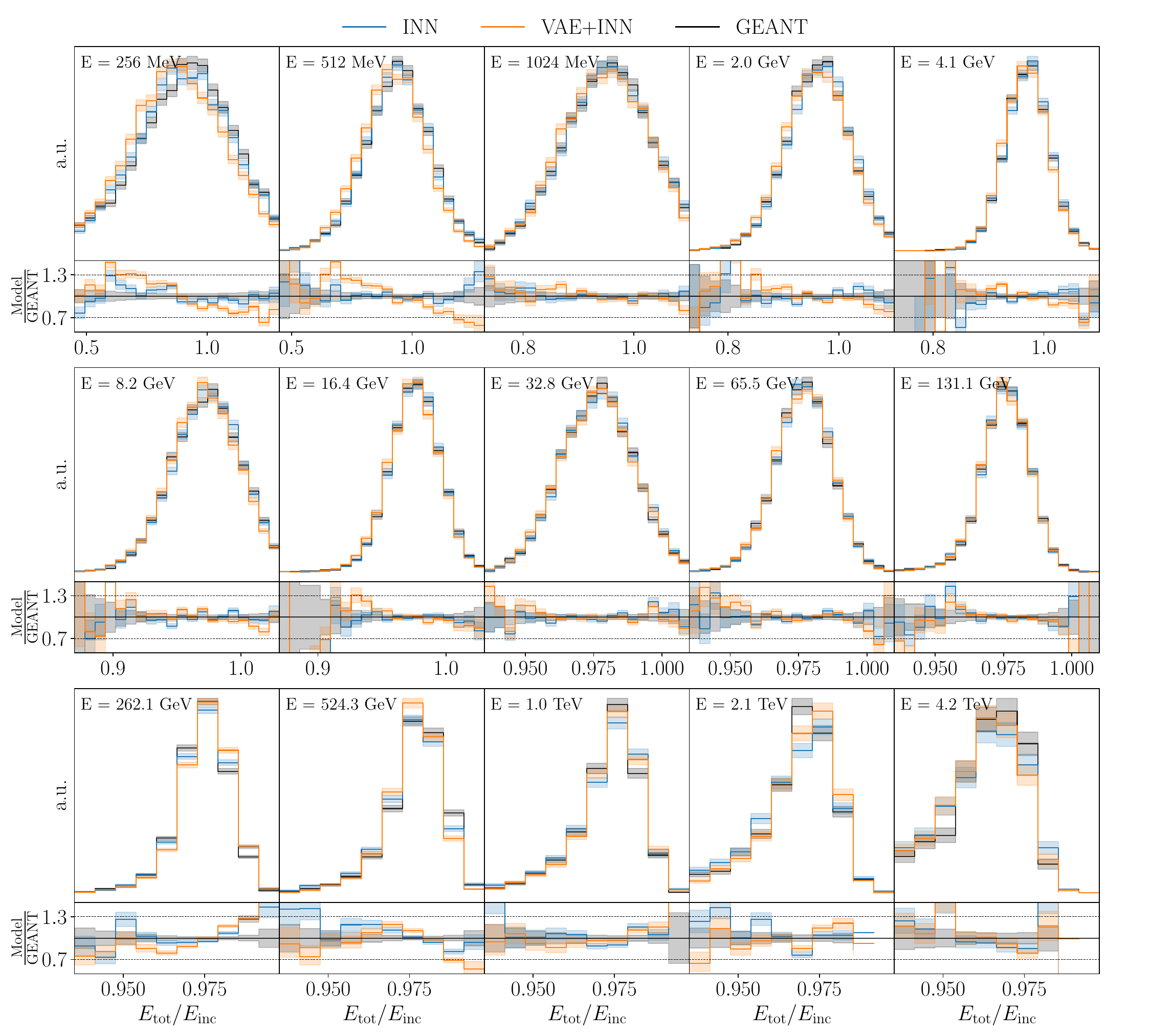}
    \caption{Energy ratio $E_\text{tot}/E_\text{inc}$ for each
      discrete incident energy, compared between \geant, INN, and
      VAE+INN for $\gamma$ showers.  }
    \label{fig:atlas-style-photon}
\end{figure}

We start with the photons in dataset~1. At high energy the interactions with matter in a photon shower are dominated by pair production and Bremsstrahlung, making this dataset the simplest in terms of dimensionality and complexity. We summarize the most
interesting high-level features for the \geant training data, the INN
generator, and the VAE+INN generator in
Fig.~\ref{fig:hlf_ds1_photons}.

We first look at the shower shape in rapidity for the layer with the largest energy deposition for $E_\text{inc} = 0.256, 8.2, 262.1$ GeV.
These energies provide insights on the generation over the
entire spectrum of low, medium, and high energetic showers.
For instance, we show the center of energy and its width in the 
calorimeter layer-1 for $E_\text{inc}=256$ MeV, while, for the remaining
energies, we show the layer~2.
Dataset~1 is not symmetric in $\eta$ and $\phi$,
because the shower were not generated around $\eta=\phi=0$.  All
showers have the same mean width, regardless of the incident
energy. This is captured by both networks at the level of 5\% to 20\%.
A failure mode of the INN is the region $\sigma_{\langle\eta \rangle}
< 20$~mm for low energies, where the network undershoots the training data by up to
50\% in the first bin.
A peculiar feature of these distributions is a small peak at
zero, which occurs when at most one voxel per layer receives a
hit. These cases are better reproduced by the VAE, whereas the INN
tends to produce slightly more collimated showers.
The INN is able to reproduce the collimated showers at higher energies 
always within the statistical uncertainties of the training data, 
both in the center of energies and their widths.

Next, we show the energy depositions in layers~1. Although the energy
deposited in layer-2 is larger for the intermediate and the large
incident energies, we focus on layer-1 to showcase the performance over a larger range of energies.
Both networks show comparable performance over the entire energy range 
for $E_\text{inc} = 8.2, 262.1$ GeV, while the VAE has larger 
deviations at lower energies.

Finally, we look at observables inclusive in the energy.
The sparsity $\lambda_2$ in the same layer is determined by the energy
threshold of 1~MeV.  The INN matches the truth over the entire
$\lambda$-range to 10\%, while the VAE struggles.  In particular, its
showers have too many active voxels, leading to the mis-modeled peak
close to zero.

The ratio $E_\text{tot}/E_\text{inc}$ exhibits a small bias
in the energy generation for the VAE+INN towards low energies,
artifact of the final threshold in the architecture. For smaller
incident energies, more voxels are zero~\cite{Diefenbacher:2023vsw},
which causes a problem for the VAE+INN because the showers are more
sparse which is the weakness of the VAE.\medskip

To illustrate the discrete structure of the incident energies in
dataset~1, we collect $E_\text{tot}/E_\text{inc}$ for each incident
energy in Fig.~\ref{fig:atlas-style-photon}. The incident energy,
provided during training and generation, carries energy-dependent
information about the shower.  For instance, low-energy showers have a
much broader energy ratio distribution, in contrast to high-energy
showers.  Both generative networks learn the conditional distribution on
$E_\text{inc}$ with deviations up to 30\% in the tails.
We include a set of shower shape histograms inclusive in the energy
in App.~\ref{app:inc_hists} and the full set of histograms for each
energy with the published samples~\cite{favaro_2024_14178546}.

\subsection{Dataset 1 pions}

\begin{figure}[b!]
    \includegraphics[width=0.33\textwidth, page=1]{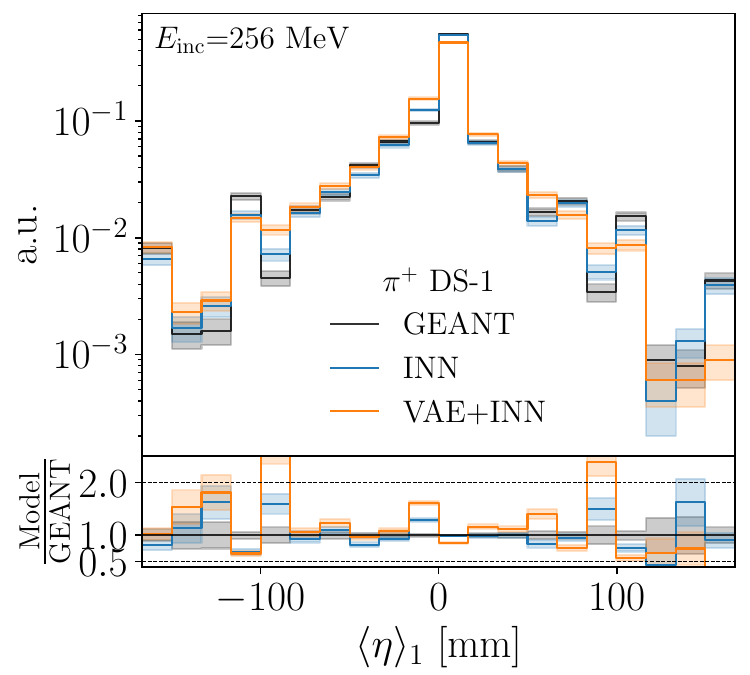}
    \includegraphics[width=0.33\textwidth, page=3]{figs/ds1-pions/E_8192/ECEta_layer_dataset_1-pions.pdf}
    \includegraphics[width=0.33\textwidth, page=4]{figs/ds1-pions/E_262144/ECEta_layer_dataset_1-pions.pdf} \\
    \includegraphics[width=0.33\textwidth, page=1]{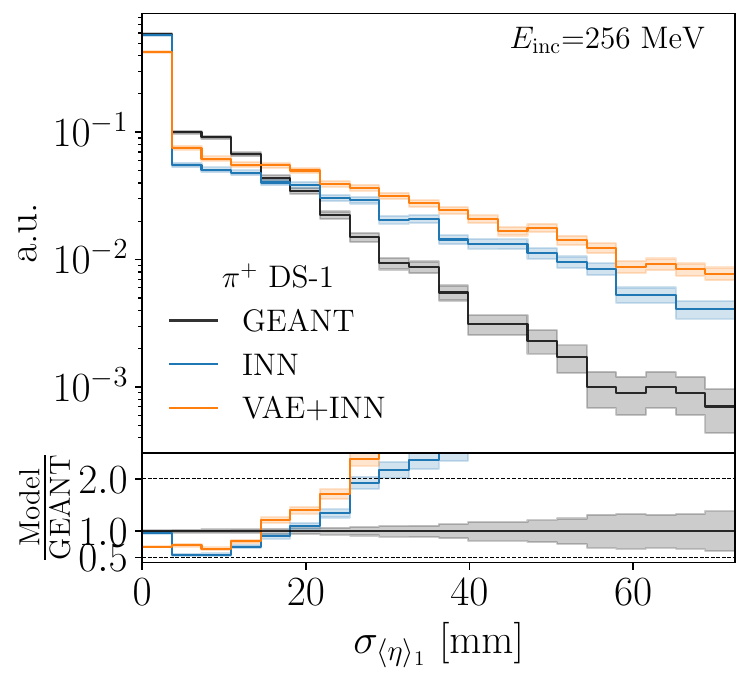}
    \includegraphics[width=0.33\textwidth, page=3]{figs/ds1-pions/E_8192/WidthEta_layer_dataset_1-pions.pdf}
    \includegraphics[width=0.33\textwidth, page=4]{figs/ds1-pions/E_262144/WidthEta_layer_dataset_1-pions.pdf} \\
    \includegraphics[width=0.33\textwidth, page=2]{figs/ds1-pions/E_256/E_layer_dataset_1-pions.pdf}
    \includegraphics[width=0.33\textwidth, page=5]{figs/ds1-pions/E_8192/E_layer_dataset_1-pions.pdf}
    \includegraphics[width=0.33\textwidth, page=6]{figs/ds1-pions/E_262144/E_layer_dataset_1-pions.pdf} \\
    \includegraphics[width=0.33\textwidth]{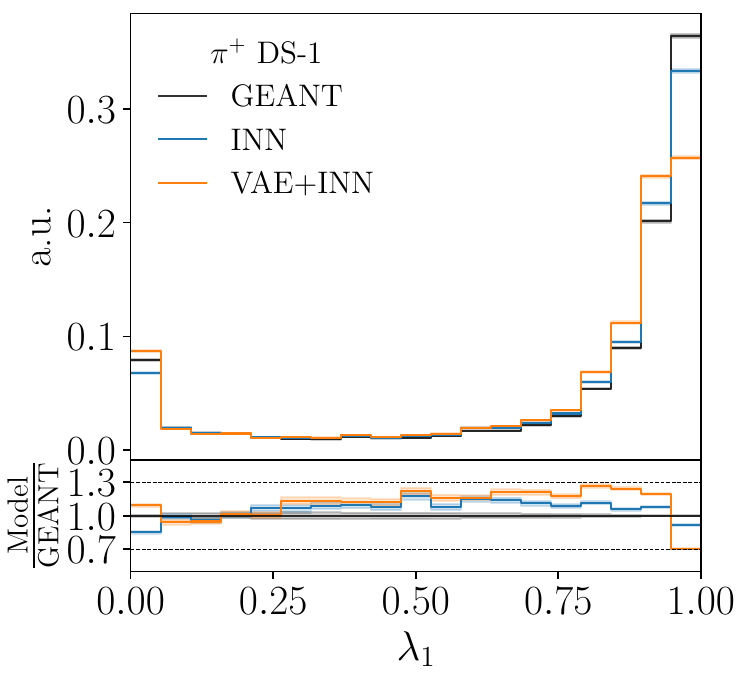}
    \includegraphics[width=0.33\textwidth]{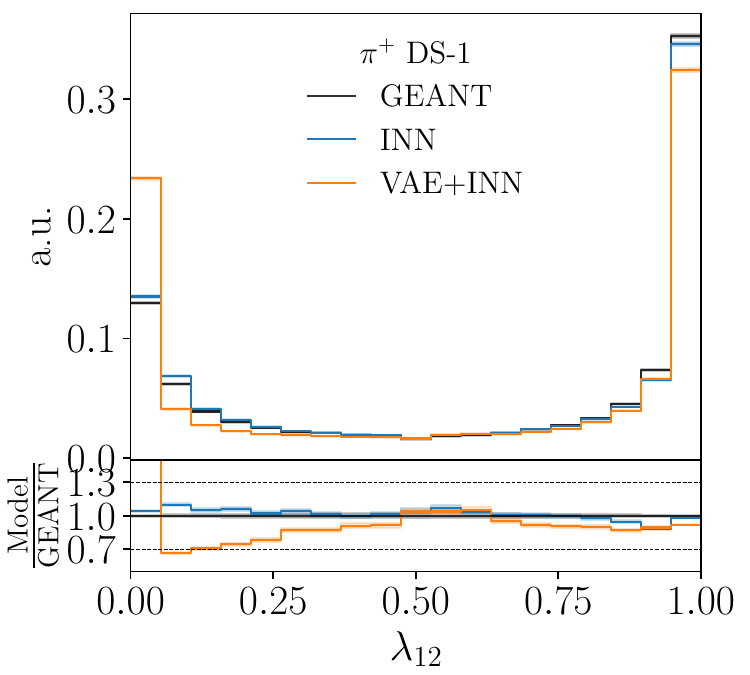}
    \includegraphics[width=0.33\textwidth]{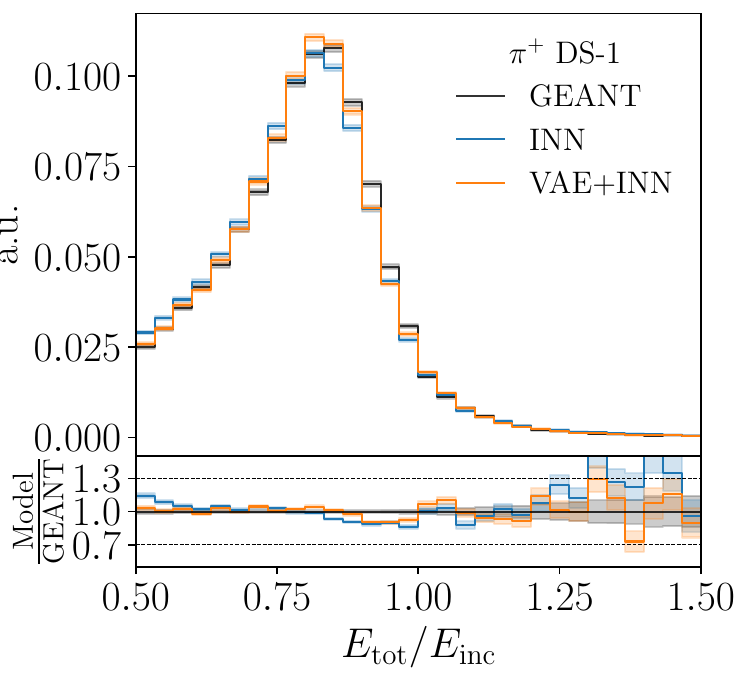}
    \caption{Set of high-level features for pion showers in dataset~1,
      compared between \geant, INN, and VAE+INN.
      We show the energy deposition, the center of energy, and the 
      width of the center of energy in layer-1, layer-12, and layer-13. For each layer, we show a single incident energy. The last row contains the inclusive sparsity in layer-1 and layer-12, and the inclusive energy ratio $E_\text{tot}/E_\text{inc}$.}
    \label{fig:hlf_ds1_pions}
\end{figure}

\begin{figure}[t]
    \centering
    \includegraphics[width=0.33\textwidth]{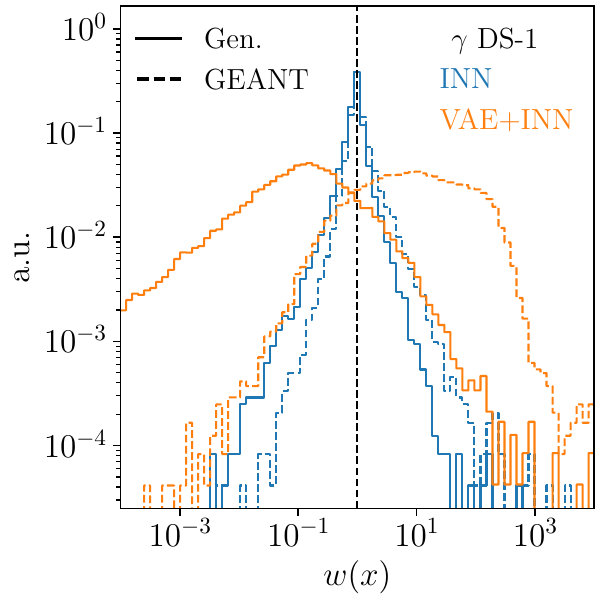}
    \hspace*{0.1\textwidth}
    \includegraphics[width=0.33\textwidth]{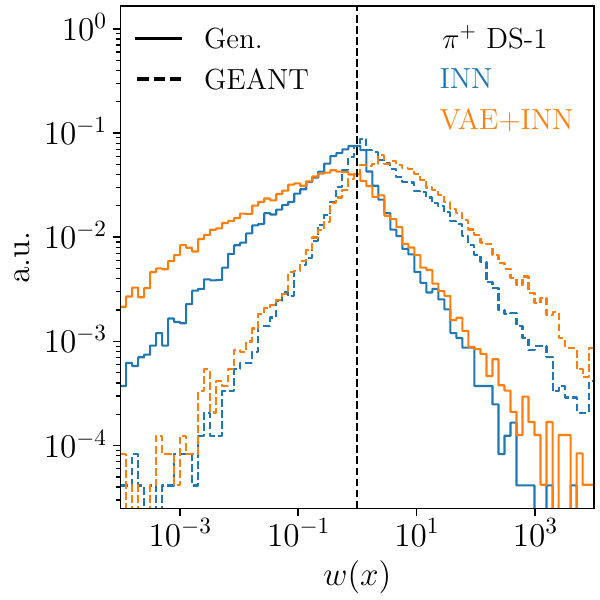}
    \caption{Classifier weight distributions in dataset 1. Classifier
      trained on low-level features for $\gamma$ showers (left) and $\pi$ showers (right). }
    \label{fig:cls_ds1}
\end{figure}

The physics of hadronic showers is significantly more complex than
photon showers, so it is interesting to see how our INNs perform for a
low-dimensional calorimeter simulation of pions.  As before, we show
shower shapes, sparsity, energy depositions, and the fraction of
deposited energy in Fig.~\ref{fig:hlf_ds1_pions}.
We focus on three distinct incident energies, 256 MeV, 8.2 GeV, and 262.1 GeV.
In particular we show the layers with the largest energy deposition. For the lowest incident energy, a large fraction of energy is deposited in layer-2 while for the other two cases we show layer-12 and layer-13, respectively.

For the shower shapes, both networks show small, percent-level
deviations in the bulk of the distributions at larger energies, while a larger discrepancy is found in the low-energy regime.
In addition, the VAE+INN
is smearing out secondary peaks of the distributions. Both networks
generate slightly too wide showers. This effect is evident only at $E_\text{inc}=256$ MeV and the more physically interesting energy region is modeled within statistical uncertainties for the INN, besides very narrow showers with width of the center of energy close to zero.

In the energy distributions we see the benefit of a smaller latent space.
The energy variables of the VAE+INN are modeled within statistical uncertainties in layer-12 and layer-13 while the INN shows a few bins with deviations up to 30\%.
Additionally, the sharp cut at low energy is smeared to a different extent by the networks.
Finally, we show global sparsity features and the $E_\text{tot}/E_\text{inc}$ ratio.
As before the VAE+INN is unable to capture the correct number of active voxels.
From the ratio $E_{\text{tot}}/E_{\text{inc}}$ we see that at all
energies the fraction of deposited energy can be very different from
shower to shower, leading to the wide energy distribution far from
one.
We collect in App.~\ref{app:inc_hists} a set of inclusive histograms and the $E_\text{tot}/E_\text{inc}$ ratio for single incident energies.

\subsubsection*{Low-level classifier}

To evaluate the performance of our generative networks on dataset~1
systematically, we train a network to learn the classifier weights
defined in Eq.\eqref{eq:def_weights} over the voxel space.  In the
left panel of Fig.~\ref{fig:cls_ds1} we show the weights for the
$\gamma$-shower.  We clearly see that the INN outperforms the VAE+INN.
Its weight distribution peaks much closer to 1 and the corresponding
AUC of 0.603(2) is substantially better than the corresponding AUC of 0.937(2)
of the VAE+INN.

More importantly, the INN does not show significant tails at large or
small weights, which would indicate distinct failure modes.  The peak
of the VAE+INN, on the other hand, has moved away from 1.  The tail at
small weights indicates regions that are overpopulated by the
network. We already know that this is the case for the sparsity.
Large weights appear in phase space regions which the VAE+INN fails to
populate, for instance the widths of the centers of energy.

In the right panel of Fig.~\ref{fig:cls_ds1} we see that the two
generators perform more similar for $\pi$-showers. Both networks now
show tails at small and large weights, two orders of magnitude away
from one. This means there are regions that are over- and
underpopulated by the generative networks. The fact that small weights
appear for generated showers and large weights appear for the training
data is generally expected.  The AUCs for the INN and the VAE+INN are
0.804(2) and 0.864(2), respectively.  The INN weight distribution is sharper
around one, resulting in the smaller AUC compared to the combined
VAE+INN approach. Altogether, we find that for dataset~1 with its
limited dimensionality of 368 voxels for photons and 533 voxels for
pions the INN works well, and that adding a VAE hinders the generative model because of a more complex latent space and a limited reconstruction quality of the autoencoder.

\subsection{Dataset 2 electrons}

\begin{figure}[htbp]
    \includegraphics[width=0.33\textwidth, page=2]{figs/ds2/E_1_10/ECEta_layer_dataset_2.pdf}
    \includegraphics[width=0.33\textwidth, page=11]{figs/ds2/E_10_100/ECEta_layer_dataset_2.pdf}
    \includegraphics[width=0.33\textwidth, page=21]{figs/ds2/E_100_1000/ECEta_layer_dataset_2.pdf} \\
    \includegraphics[width=0.33\textwidth, page=2]{figs/ds2/E_1_10/WidthEta_layer_dataset_2.pdf}
    \includegraphics[width=0.33\textwidth, page=11]{figs/ds2/E_10_100/WidthEta_layer_dataset_2.pdf}
    \includegraphics[width=0.33\textwidth, page=21]{figs/ds2/E_100_1000/WidthEta_layer_dataset_2.pdf} \\
    \includegraphics[width=0.33\textwidth, page=2]{figs/ds2/E_1_10/E_layer_dataset_2.pdf}
    \includegraphics[width=0.33\textwidth, page=11]{figs/ds2/E_10_100/E_layer_dataset_2.pdf}
    \includegraphics[width=0.33\textwidth, page=21]{figs/ds2/E_100_1000/E_layer_dataset_2.pdf}
    \includegraphics[width=0.33\textwidth, page=11]{figs/ds2/Sparsity_layer_dataset_2.pdf} 
    \includegraphics[width=0.33\textwidth, page=21]{figs/ds2/Sparsity_layer_dataset_2.pdf} 
    \includegraphics[width=0.33\textwidth]{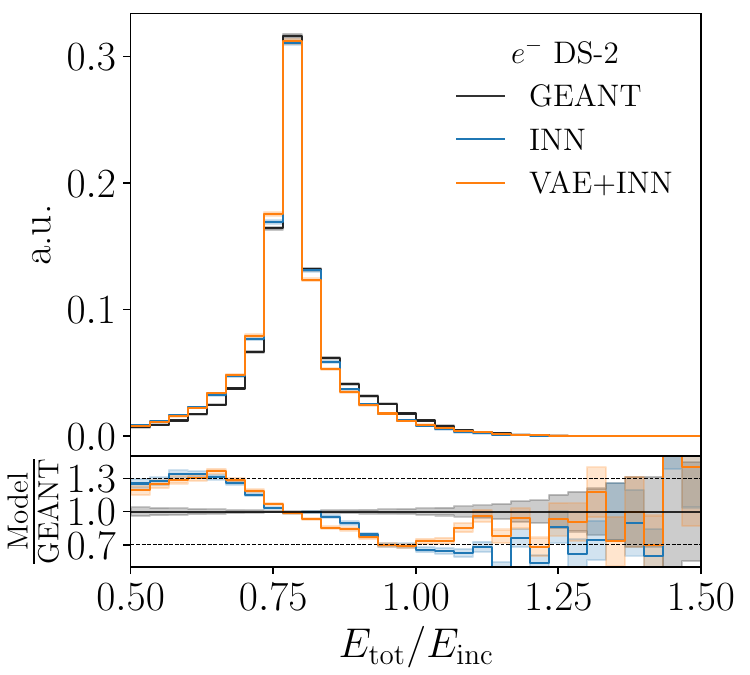}
    \caption{Set of high-level features for electron showers in
      dataset~2, compared between \geant, INN, and VAE+INN.
      We show the energy deposition, the center of energy, and the 
      width of the center of energy in layer-1, layer-10, and layer-20. For each layer, we show a single incident energy range. The last row contains the inclusive sparsity in layer-10 and layer-20, and the inclusive energy ratio $E_\text{tot}/E_\text{inc}$.}
    \label{fig:hlf_ds2}
\end{figure}

Dataset~2 is given in terms of 6480 voxels, the kind of dimensionality
which will probe the limitations of the regular INN. The number of
parameters for this network approaches 200M.  The question will be, if
the VAE+INN condensation helps the performance of the network.  As
before, we show a representative set of high-level features in
Fig.~\ref{fig:hlf_ds2}. This time we group the showers in three equally spaced energy windows in $\log E_{\text{inc}}$. For brevity, we only focus on three different layers representative for the interaction of the incident particle with the detector in each energy window. We include all the remaining histograms together with the published samples.
We choose layer~1 for $E_{\text{inc}} \in [10^0, 10^1]$ GeV, layer-10 for $E_{\text{inc}} \in [10^1, 10^2]$ GeV, and layer-20 for showers with incident energy in $E_{\text{inc}} \in [10^2, 10^3]$ GeV.

From the shower shapes we see that the INN-based architecture
generates realistic showers at all energies. The training is stable and
consistent across different runs of the same architecture. We observe agreement in the center of energy distributions with small deviations towards larger $\langle \eta \rangle$ and less commonly around zero, as shown in layer-10.
The agreement in phase space density between \geant and
the INN ranges from a few percent in the bulk of the distributions to
50\% in the tails, where very low statistics is available. Similar numbers apply to the width of the center of
energy a shift towards wider showers is observed in all the energy windows. The failure mode of the INN, regardless of the dataset, is an
under-sampling of showers with width between the peak at zero and the
secondary peak, for which the location depends on the layer but not on
the incident energy. 

The VAE+INN shows limitations for large incident energies. As the energy increases, the generated showers only reproduce the mean value of the center of energy, i.e.~showers have a rather uniform energy distribution around the center of the calorimeter resulting in $\langle \eta \rangle$ peaked around zero.
A similar failure mode is observed in the width of the center of energy where the VAE+INN is more concentrated around the mean width of the \geant showers.
On the other hand, the compression mechanism of the VAE works well at lower energies where showers are generated by the latent INN and reconstructed by the VAE within statistical precision besides the $\sigma_{\langle \eta \rangle}$ region close to zero.

The two networks learn the energy depositions in the layers in two
very different spaces. The INN extracts them with a large number of
voxels, while the VAE+INN compresses them into a reduced space of
around 50 features. This different dimensionality is
reflected in all energy distributions in Fig.~\ref{fig:hlf_ds2}.
While in poorly populated tails the INN does slightly better, the VAE+INN performs better for the main features in the
central and high-energy regime. This is true for the layer-wise
energies, but also for the ratio $E_\text{tot}/E_\text{inc}$.
The last two plots show the inclusive sparsity distributions in layer-10 and layer-20 confirming that the INN reproduces the sparsity across the entire phase space while the VAE+INN struggles from the decoding step.

\subsubsection*{Low-level classifier}

\begin{figure}[t]
    \centering
    \includegraphics[width=0.33\textwidth]{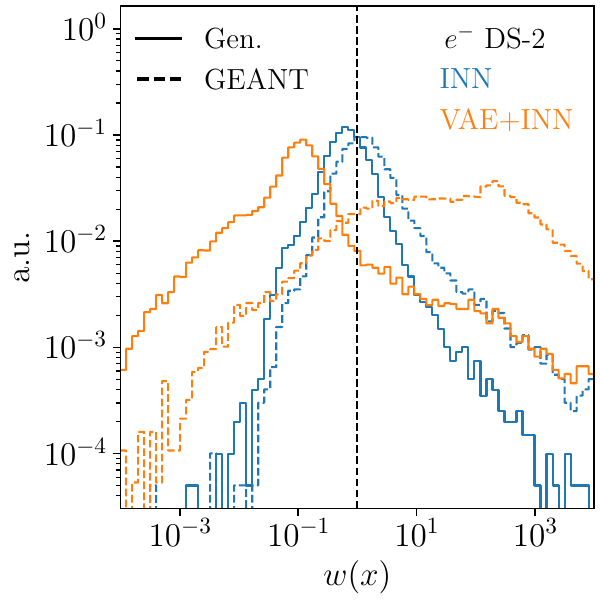}
    \hspace*{0.1\textwidth}
    \includegraphics[width=0.33\textwidth]{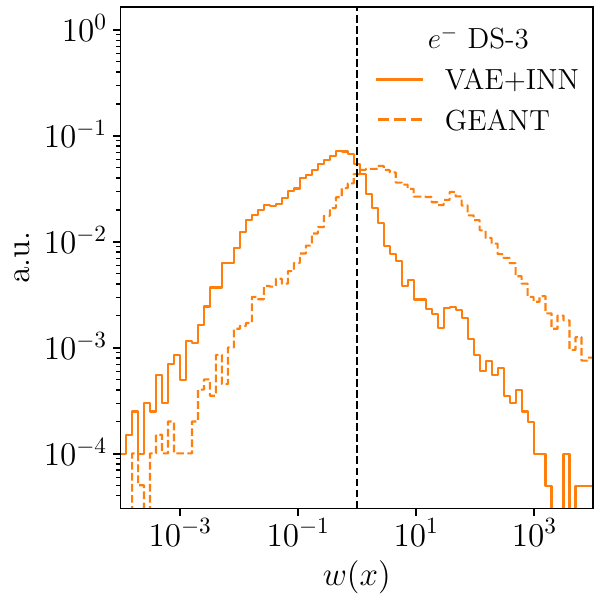}
    \caption{Classifier weight distributions. Classifier trained on
      $e^{-}$ showers on dataset 2 (left) and dataset 3 (right). The
      tails of dataset~3 should be taken with a grain of salt, giving
      the limitations of the simple classifier architecture.}
    \label{fig:cls_ds23}
\end{figure}

Again, we show a systematic comparison for dataset~2 in terms of the
classifier weights in the left panel Fig.~\ref{fig:cls_ds23}.
Compared to dataset~1, there is a clear deterioration of the INN
performance for the higher-dimensional phase space. At small weights,
the tail remains narrow, indicating that there are still no phase
space regions where the network over-samples the true phase space
distribution. For large weights the weight tail now extends to values
larger than $w \sim 10^3$. This tail can be related to a recurrent
under-sampling of showers with a small width of the center of energy
in each layer, as seen in Fig.~\ref{fig:hlf_ds2}.

The classifier evaluating the VAE+INN generator highlights a few
important structures as well.  First, we have a clear over-sampled
region in phase space with weights $w \sim 10^{-2}$, which we can
relate to the center of energy distribution as well. As mentioned
before, the VAE+INN over-samples showers with width close to the mean
shower width. The classifier confirms this major failure mode. For the
large-weight tail we checked that the under-sampled showers do not
cluster in the same way, but are distributed over phase space,
including tails of distributions.

The AUC values of the classifiers for dataset~2, 0.705(5) for the INN and
0.916(3) for the VAE+INN, confirm the challenge of the INN related to the size of the model and the dimensionality of dataset-2, especially
relative to the well-modeled $\gamma$-showers in dataset~1.
Adding a VAE improves the generation of low-energetic showers, where the low activity inside the calorimeter can be nicely compressed in the latent space, however this is out-weighted by the failure modes observed at medium and high incident energies.

\subsection{Dataset 3 electrons}

\begin{figure}[t]
    \includegraphics[width=0.33\textwidth, page=21]{figs/ds3/ECEta_layer_dataset_3.pdf}
    \includegraphics[width=0.33\textwidth, page=21]{figs/ds3/WidthEta_layer_dataset_3.pdf}
    \includegraphics[width=0.33\textwidth, page=21]{figs/ds3/Sparsity_layer_dataset_3.pdf} \\
    \includegraphics[width=0.33\textwidth, page=1]{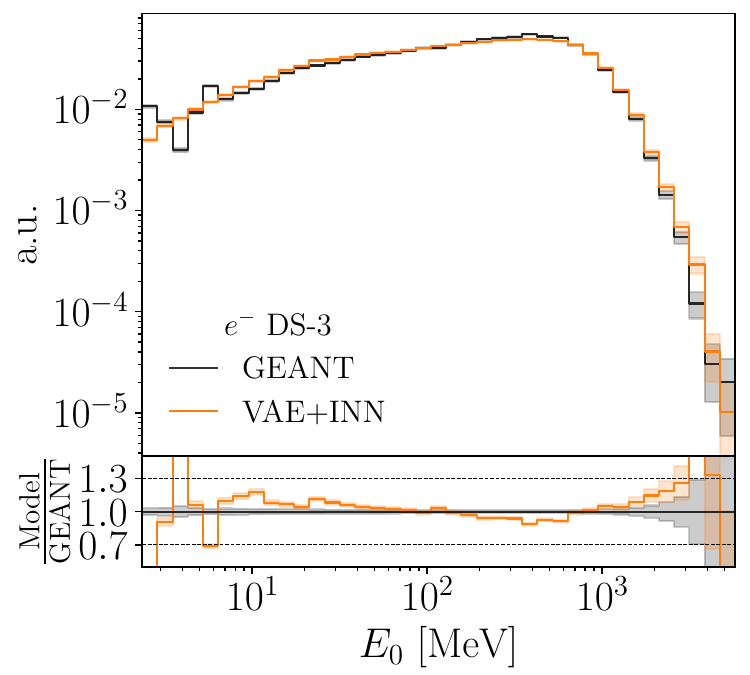}
    \includegraphics[width=0.33\textwidth, page=21]{figs/ds3/E_layer_dataset_3.pdf}
    \includegraphics[width=0.33\textwidth]{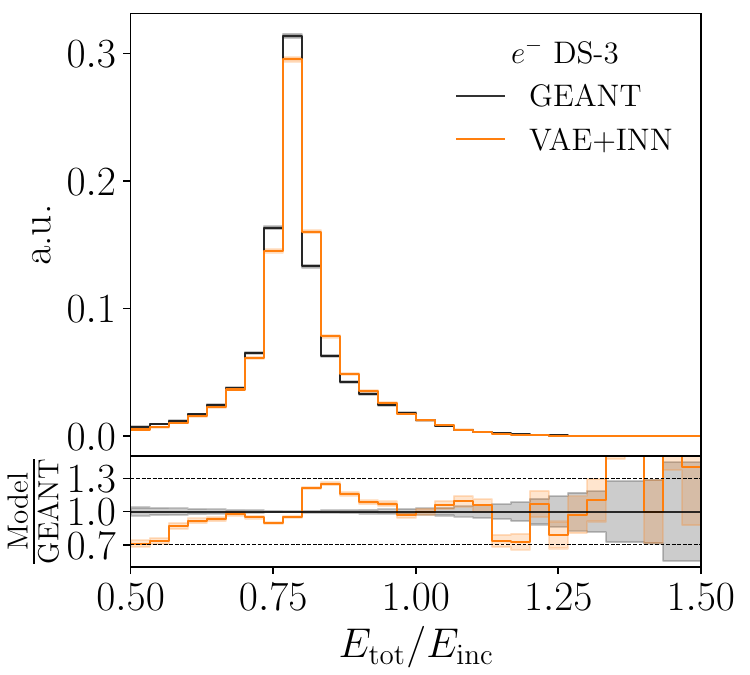}
    \caption{Set of high-level features for electron showers in
      dataset~3, compared between \geant, INN, and VAE+INN.}
    \label{fig:hlf_ds3}
\end{figure}

Finally, we tackle dataset~3, which includes the same physics as
dataset~2, but over a much higher-dimensional and extremely sparsely
populated phase space.  For this dataset we cannot train an INN
without dimensionality reduction, so we only show VAE+INN results in
Fig.~\ref{fig:hlf_ds3}. As expected, the performance is worse than for
dataset~2, but the training is stable across different training
runs. One problem is a worsening reconstructions of the centroids and
widths for the later layers, which is likely related to the small
average energy deposition per voxel. The maximum in the width
distributions is overpopulated by the VAE+INN. For the
energy distributions the VAE+INN is doing reasonably well, with
serious deviations only in the low-energy tails.

\begin{table}[b!]
\centering
\begin{small} \begin{tabular}{l|l|@{\hskip 10pt}r@{\ $\pm$ \ }l@{\hskip 10pt}r@{\ $\pm$ \ }l@{\hskip 10pt}r@{\ $\pm$ \ }l@{\hskip 10pt}r@{\ $\pm$ \ }l} \toprule
\multirow{2}{*}{}   & \multirow{2}{*}{Batch size} & \multicolumn{8}{c}{INN} \\ 
                     &         & \multicolumn{2}{c}{1-photon}   & \multicolumn{2}{c}{1-pion}   & \multicolumn{2}{c}{2-electron}   & \multicolumn{2}{c}{3-electron}   \\ \midrule
\multirow{3}{*}{GPU} & 1       & 24.79 & 0.49  & 24.76 & 0.35  & 50.90 & 0.37  & \multicolumn{2}{c}{} \\
                     & 100     & 0.385 & 0.002 & 0.406 & 0.003 & 1.900 & 0.026 & \multicolumn{2}{c}{} \\
                     & 10000   & 0.162 & 0.002 & 0.191 & 0.006 &  \multicolumn{2}{c}{exceeding memory}   & \multicolumn{2}{c}{} \\ \midrule
\multirow{3}{*}{CPU} & 1       & 17.48 & 0.09  & 18.88 & 0.33  & 117.5 & 1.8   & \multicolumn{2}{c}{} \\
                     & 100     & 0.827 & 0.028 & 1.004 & 0.047 & 14.26 & 0.18  & \multicolumn{2}{c}{} \\
                     & 10000   & 0.510 & 0.008 & 0.719 & 0.016 & 15.24 & 1.36  & \multicolumn{2}{c}{} \\ \toprule
\multirow{2}{*}{}    & \multirow{2}{*}{Batch size} & \multicolumn{8}{c}{VAE+INN} \\ 
                     &         & \multicolumn{2}{c}{1-photon}   & \multicolumn{2}{c}{1-pion}   & \multicolumn{2}{c}{2-electron}   & \multicolumn{2}{c}{3-electron} \\ \midrule
\multirow{3}{*}{GPU} & 1       & 33.64 & 0.32  & 33.54 & 0.23  & 40.55 & 0.40  & 43.13 & 1.4 $^*$  \\
                     & 100     & 0.507 & 0.005 & 0.544 & 0.007 & 1.05  & 0.02  & 3.44  & 0.04    \\
                     & 10000   & 0.180 & 0.002 & 0.228 & 0.003 & 0.748 & 0.018 & \multicolumn{2}{c}{---}  \\ \midrule
\multirow{3}{*}{CPU} & 1       & 20.83 & 0.72  & 20.05 & 0.13  & 28.11 & 0.15  & 39.46 & 1.1 $^*$   \\
                     & 100     & 0.582 & 0.005 & 0.886 & 0.015 & 1.94  & 0.01  & 4.91  & 0.01    \\
                     & 10000   & 0.328 & 0.004 & 0.426 & 0.014 & 1.25  & 0.01  & 4.97  & 0.08   \\ \bottomrule
\end{tabular} \end{small}
\caption{Per-shower generation times in ms. We show mean and standard
  deviation of 10 independent runs. The star indicates that only 10k
  samples were generated. The CPU timings were done with an Intel(R)
  Core(TM) i9-7900X at 3.30~GHz, the GPU timings with an NVIDIA
  TITAN~V with 12GB RAM.}
\label{tab:gen_timeit}
\end{table}

The classifier weights for the VAE+INN generating dataset~3 are shown
in the right panel of Fig.~\ref{fig:cls_ds23}.  Even though the
generative task is considerably harder, the learned weight
distribution broadens centrally, but shows smaller tails than for
dataset~2. The reason is that not only the generative network, but
also our simple classifier are reaching their limits. However, the
bulk of the classifier weight distribution clearly indicates that for
dataset~3 the phase space density is mis-modelled by factors as large
as 100 or 0.01 over large phase space regions. While the INN
description of these position showers fails altogether, the VAE+INN
results do not guarantee the level of precision we would expect for
generative networks at the LHC.
For completeness, we include the high-level distributions divided in windows of incident energies only with the published samples~\cite{favaro_2024_14178546}.

\subsection{Comparison}

\begin{table}[]
\centering
\begin{small}
\resizebox{0.9\textwidth}{!}{
    \begin{tabular}{c|@{\hskip 5pt}r@{\ / \ }l@{\hskip 5pt}r@{\ / \ }l@{\hskip 5pt}r@{\ / \ }l}
     \toprule 
      & \multicolumn{6}{c}{Generation time in ms per shower (CPU/GPU)} \\
      & \multicolumn{6}{c}{Batch size 1} \\
         & \multicolumn{2}{c}{1-photon} & \multicolumn{2}{c}{1-pion} & \multicolumn{2}{c}{2-electron}  \\ \midrule
         CaloINN & 38(3) & 25(2) & 43(3) & 25(2) & 3.9(3)$\cdot 10^2$ & 53(1) \\
         CaloFlow teacher\cite{Krause:2022jna} & 4.3(3)$\cdot 10^4$ & 4.2(1)$\cdot 10^3$ & 2.0(3)$\cdot 10^5$ & 6.2(1)$\cdot 10^3$ & \multicolumn{2}{c}{---} \\
         CaloFlow student\cite{Krause:2022jna} & 5.7(2)$\cdot 10^2$ & 56.9(5) & 6.2(2)$\cdot 10^2$ & 77(4) & \multicolumn{2}{c}{---} \\
         CaloDiffusion\cite{Amram:2023onf} & 1.57(6)$\cdot 10^4$ & 5.59(6)$\cdot 10^3$ & 1.5(1)$\cdot 10^4$ & 5.67(5)$\cdot 10^3$ & 3.6(2)$\cdot 10^4$ & 5.29(8)$\cdot 10^3$ \\ \midrule
         
         & \multicolumn{6}{c}{Batch size 100} \\
         & \multicolumn{2}{c}{1-photon} & \multicolumn{2}{c}{1-pion} & \multicolumn{2}{c}{2-electron}  \\ \midrule
         CaloINN & 2.7(3) & 0.51(3) & 3.9(4) & 0.44(1) & 60(10) & 1.18(3) \\
         CaloFlow teacher\cite{Krause:2022jna} & 2.0(1)$\cdot 10^3$ & 45(1) & 5.4(5)$\cdot 10^3$ & 70(1) & \multicolumn{2}{c}{---} \\
         CaloFlow student\cite{Krause:2022jna} & 11(1) & 0.79(1) & 14(2) & 1.00(2) & \multicolumn{2}{c}{---} \\
         CaloDiffusion\cite{Amram:2023onf} & 4.6(3)$\cdot 10^3$ & 75(2) & 1.57(6)$\cdot 10^4$ &77(2) & 2.3(3)$\cdot 10^4$ & 99(2) \\ \bottomrule
    \end{tabular}
    }
    \caption{Generation time for networks trained on the full space. We compare our network CaloINN, CaloFlow teacher and student, and CaloDiffusion. Numbers are taken from the CaloChallenge~\cite{Krause:2024avx} review.}
    \label{tab:time_comp}
\end{small}
\end{table}

\begin{table}[]
\centering
\begin{small}
    \begin{tabular}{c|@{\hskip 10pt}r@{\ / \ }l@{\hskip 10pt}r@{\ / \ }l@{\hskip 10pt}r@{\ / \ }l}
     \toprule 
      & \multicolumn{6}{c}{AUC (LL/HL)} \\
        & \multicolumn{2}{c}{1-photon} & \multicolumn{2}{c}{1-pion} & \multicolumn{2}{c}{2-electron}  \\ \midrule
         CaloINN & 0.603(2) & 0.563(3) & 0.804(2) & 0.692(1) & 0.705(5) & 0.891(2) \\
         CaloFlow teacher\cite{Krause:2022jna} & 0.701(3) & 0.551(3) & 0.827(3) & 0.692(2) & \multicolumn{2}{c}{---} \\
         CaloDiffusion\cite{Amram:2023onf} & 0.62(1) & 0.62(1) & 0.65(1) & 0.65(1) & 0.56(1) & 0.56(1) \\ \bottomrule
    \end{tabular}
    \caption{Summary of low-level (LL) and high-level (HL) classifier scores for our networks, where errors are extracted from 10 different classifiers. We compare to CaloFlow\cite{Krause:2022jna} and CaloDiffusion\cite{Amram:2023onf}. An in-depth comparison between different architectures, including latent models, is part of \cite{Krause:2024avx}.}
    \label{tab:auc_comp}
\end{small}
\end{table}

Finally, we compare our INN to other networks in two main aspects, the generation timing and the shower fidelity as measured by the AUC of a classifier. First, we do an in-depth timing study of our networks using the CaloChallenge~\cite{Krause:2024avx} procedure.  The
INN architecture with modern coupling layers is ideally suited for
fast and precise generation.  We create a singularity
container~\cite{10.1371/journal.pone.0177459} of the software
environment and take the time it takes to load the container, load the
network, move it on the GPU, generate the samples, and save them to
disk.  In Tab.~\ref{tab:gen_timeit} we show the averaged results from
ten runs. We observe a speed--up for increased batch size and when
running on the GPU. The INN has a small advantage for dataset~1, but
is unable to generate dataset~2 with the highest batch size and
dataset~3 altogether. The VAE shows generation times at or below the
millisecond mark.

The training time for the DS1 network on a single A30 GPU is $\sim$ 4 hours, including the validation steps which slightly increased the training time due to the large number of validation figures. However, we are unable to provide an exact number because of large fluctuations in the training time coming from the shared CPU and GPU memory of the cluster.
Under the same observations, the network for dataset-2 trained on average for $\sim$ 6 hours.

The generation time from different published networks can vary because of varying hardware, making a fair comparison laborious.
To avoid generating samples from different networks,
we base our timing comparison on the result of the CaloChallenge~\cite{Krause:2024avx}.
We look at two well-known architectures based on autoregressive normalizing flows, CaloFlow teacher and student~\cite{Krause:2022jna}, and the diffusion model CaloDiffusion~\cite{Amram:2023onf}, which provides benchmark results for this class of neural networks.
From Tab.~\ref{tab:time_comp}, we observe that the coupling block structure provides a generation speed-up when compared to the autoregressive counterpart, as studied in the CaloFlow architecture.
As expected, our model is also substantially faster than a diffusion model due to the additional function evaluation needed to revert the diffusion process.

We summarize the shower fidelity results in Tab.~\ref{tab:auc_comp}. Our figure of merits
are a classifiers trained on all the voxels, i.e. low-level features, and a second one trained
on high-level observables. These include the layer energy deposition, the center of energy and the width of the center of energy in both $(\eta, \phi)$ directions, and the incident energy.
From the AUC score, our photon showers on dataset-1 show the best performance on the low-level feature, even when compared to current diffusion networks, and high-level features comparable to CaloFlow.
For pions, the complex low-level shower structure is better captured by CaloDiffusion, while flow networks still retain good high-level shower quality.
Lastly, the training challanges encountered while training on dataset-2 are also reflected on the ability of the CaloINN network
to generate high-dimensional electron showers.

\section{Conclusions}
\label{sec:conclusion}

Simulations are at the heart of the LHC program. Modern generative
networks are showing great promise to improve their quality and speed,
allowing them to meet the requirements of the high-luminosity LHC. In
this paper, we have studied fast and precise normalizing flows,
specifically an INN and a VAE+INN combination to generate calorimeter
showers in high-dimensional phase spaces. As reference datasets we use
the CaloChallenge datasets~1 to~3, with an increasing number
of 368, 533, 6480, and 40,500 voxels.

For the simplest dataset~1 photon showers, we have found that the
INN generated high-fidelity showers and learns the phase space density
of high-level features at the 10\% level, except for failure modes
which we can identify using high-level features and classifier weights
over the low-level phase space. 
We have found that the INN provides unmatched speed with, for instance, $\mathcal{O}(10)$ ms generation time on a single CPU for a single shower. At the same time the shower quality is comparable or even better than other deep generative networks, including diffusion models.

For the pions in dataset~1 the INN faces more serious challenges,
including mis-modeled features, and a wider range of learned classifier
weights. The performance difference between the INN and the VAE+INN 
is limited by the expressivity of the latent network, with the additional sparsity failure mode introduced by the autoencoder.
Also in this case our networks show shower accuracy comparable to other normalizing flows while providing faster generation time.

The electron showers in dataset~2 introduced technical challenges for the full-space INN. We have observed that the compression by the
VAE+INN helps learning simple showers in the high-dimensional calorimeter. In particular, the main shape features of low-energetic showers are improved by the VAE+INN, including all the energy variables which are learned in a smaller latent space.
However, the compression introduces more complex features for higher energies, where we observe a substantial deterioration of the VAE+INN.
The second issue is introduced by the decoding step, which limits the reconstruction of low-energy depositions and the sparsity. 
The INN produces good-quality showers across the entire phase space.
Although the physics is similar to the photons in dataset-1, we have not matched the same shower quality. This is attributed to difficulties in the optimization task of a much larger INN.
In this paper we focused on the improvements provided by an INN for fast detector simulation and we believe other architectures can also benefit from our observations. For instance, a super-resolution approach, as in~\cite{Pang:2023wfx}, would show better scaling properties to higher-dimensional calorimeters while retaining the improvement proposed by our CaloINN.
Additionally, having a separate network which learns only the energy variables will further increase the overall fidelity in terms of both layer energy deposition and shower shape observables.

Finally, the electrons in
dataset~3 exceed the power of the plain INN, leaving us with the
VAE+INN as the remaining option. As for dataset-2, we have observed an intrinsic limitation of the latent INN to learn the compressed features which we partially addressed by moving to a kernel-based autoencoder architecture. While still providing fast generation, the VAE+INN is not able to generate high-fidelity showers. Although the expressivity of the network can be improved, as it has been done in~\cite{Favaro:2024rle}, generating the correct sparsity remains an open question for latent models.

The generated samples used in this paper are published on Zenodo at \href{https://zenodo.org/records/14178546}{10.5281/zenodo.14178546}. We also include the complete set of high-level features for the studied incident energies and the inclusive results. 

\subsection*{Acknowledgements}

We would like to thank Theo Heimel, Stefan Radev and Peter Loch for 
helpful discussions. We would like to thank Thorsten Buss for collaborating in an early phase of the project. We would
like to thank the Baden-W\"urttem\-berg Stiftung for financing through
the program \textsl{Internationale Spitzenforschung}, pro\-ject
\textsl{Uncertainties – Teaching AI its Limits} (BWST\_ISF2020-010). DS
is supported by the U.S.~Department of Energy under Award Number
DOE-SC0010008. This research is supported by the Deutsche
Forschungsgemeinschaft (DFG, German Research Foundation) under grant
396021762 -- TRR~257: \textsl{Particle Physics Phenomenology after the
  Higgs Discovery} and through Germany's Excellence Strategy
EXC~2181/1 -- 390900948 (the \textsl{Heidelberg STRUCTURES Excellence
  Cluster}).

\clearpage
\appendix
\section{Appendices}
\subsection{Network details}
\label{app:hyperparameters}

In this appendix we give details on the network architectures and the 
preprocessing. The INN and the VAE+INN take voxels normalized by the
layer energy as input. The extra energy dimensions, calculated as in Eq.~\ref{eq:enc_energy},
are appended to the feature vector.

In the INN, we apply uniform noise and 
and a regularized logarithmic transformation with strength $\alpha$. 
The transformation applied to the features is a rational quadratic spline~\cite{durkan2019neural} for
dataset 1 and a cubic spline~\cite{durkan2019cubic} for dataset 2.
The prediction of the spline parameters is obtained with an MLP sub-network with 
256 nodes for each hidden layer. To equally 
learn each dimension, we permute the order of the features after a 
transformation and normalize the output to mean zero
and unit standard deviation with an ActNorm~\cite{glow} layer.
In the large-scale architecture, we stack twelve blocks to construct the INN 
with the additional preprocessing block.

The VAE preprocessing has a similar structure. After normalization, we apply
an $\alpha$-regularized logit transformation and a normalization to zero mean
and unit standard deviation to each feature. We do not add noise during training and
we set the latent dimension to 50 for dataset 1 and 2, and to 300 for dataset 3.
We provide the full list of parameters in Tabs.~\ref{tab:hyp_inn} and~\ref{tab:hyp_vae}.

The selection of the hyperparameters is based on heuristic observations from which we performed a rather limited grid search around the starting set.
For instance, the hidden dimension of the layers is chosen to be similar to the number of input features used to predict the spline parameters and the selection of the number of layers is based on previous experience on INNs~\cite{Butter:2021csz, Bellagente:2021yyh}.
While RQS are currently one of the most expressive transformations, we found occasional run-to-run instabilities while training on DS2 while the cubic spline consistently converged.
A change in batch size is accompanied by a change in the number of epochs, such that the number of iterations is approximately constant,

The initial conditions for the number of bins and the number of blocks is based on the number of parameters used in~\cite{Krause:2021ilc}, which already showed great generation performance.
Additionally we tested $b\in \{1\cdot 10^{-6}, 5\cdot 10^{-6}\, 1\cdot 10^{-5}\}$, a batch size $\in \{64, 128, 256, 512\}$, and change the number of blocks  by two units.\\
For the VAE we additionally varied the latent space by a factor of 2 up and down without any visible improvements. The size of the embedding and decoding networks was not the goal of a larger optimization. It turned out beneficial to inflate the dimensionality in the first layer in all tried configurations. Apart from that we found generally better results the larger the encoder and decoder networks were constructed. So we chose the parameter number based on the available GPU RAM.\\
We avoided large resources consumption which would be required for a finer ablation study. Therefore, we expect that the results can be improved in terms of timings, network complexity, and performance.

\begin{table}[h!]
	\centering
	\begin{small} \begin{tabular}{l@{\hskip 10pt}|@{\hskip 10pt}c@{\hskip 5pt}@{\hskip 10pt}c}
		\toprule
		Parameter          & INN DS1/DS2 & INN (with VAE)  \\
		\midrule
		coupling blocks    &   RQS / Cubic  & RQS \\
		\# layers          &    4 / 3   & 3 \\
		hidden dimension   &   256  & 32 \\
		\# of bins         &   10   & 10 \\
		\# of blocks       &   12/14   & 18 \\
		\# of epochs       &   450 / 200  & 200 \\
		batch size         &   512 / 256  &  256 \\ 
		lr scheduler       & one cycle  &  one cycle \\
		max. lr            & $1\cdot10^{-4}$  & $1    \cdot10^{-4}$ \\
		$\beta_{1,2}$ (ADAM)      & $(0.9, 0.999)$  & $(0.9, 0.999)$ \\
		$b$                & $5\cdot10^{-6}$ & / \\
		$\alpha$           & $1\cdot10^{-8}$ & $1 \cdot 10^{-6}$ \\
		\bottomrule
    \end{tabular} \end{small}
	\caption{Network and training parameters for the pure INN.}
	\label{tab:hyp_inn}
\end{table}

\begin{table}[h!]
  \centering
  \begin{small} \begin{tabular}{l@{\hskip 10pt}|@{\hskip 10pt}c@{\hskip 5pt}l}
        \toprule
		Parameter          & VAE  \\
		\midrule
		lr scheduler       &  Constant LR  & \hspace{0em}\rdelim\}{10}{*}[Inner VAE]\\
		lr                 &  $1\cdot10^{-4}$ \\
		hidden dimension   & 5000, 1000, 500 (Set 1) \\
                           & 1500, 1000, 500 (Set 2) \\
                           & 2000, 1000, 600 (Set 3) \\
        latent dimension   & 50 (Set 1,2) / 300 (Set 3) \\
        \# of epochs       & 1000 \\
		batch size         & 256 \\ 
        $\beta$            & $1 \cdot 10^{-9}$\\
        threshold $t$ [keV]& 2 (Set 1) / 15.15 (Set 2,3)\\
        \midrule
		hidden dimension   & 1500, 800, 300  & \hspace{0em}\rdelim\}{3}{*}[Kernel]\\
        kernel size        & 7 \\
        kernel stride      & 3 (Set 2), 5 (Set 3)\\
		\bottomrule
	\end{tabular} \end{small}
  \caption{Network and training parameters for the VAE-INN.}
  \label{tab:hyp_vae}
\end{table}


The classifiers trained for the evaluation of the generative networks
are simple MLP networks with leaky ReLU.  We use three
layers with 512 nodes each and a batch size of 1000.  The network is
trained for 200 epochs with a learning rate of $2\cdot10^{-4}$ and the
Adam optimizer with standard parameters. To prevent overfitting,
especially for the larger datasets, we apply 30\% dropout to each
layer, and we reduce the learning rate on plateau with a decay factor
of 0.1 and decay patience of 10.  The splitting between training,
validation, and testing is 60/20/20\%.  The selection of the best
network is based on the best validation loss.

\subsection{CaloGAN dataset}
\label{app:calogan}

In this section we discuss the INN performance on the even 
simpler CaloGAN dataset~\cite{Paganini:2017hrr,Paganini:2017dwg}.
The INN architecture is 
described in Sec.~\ref{sec:inn}. To extract uncertainties from 
the generative network, we promote the deterministic INN to its 
Bayesian counterpart~\cite{Bellagente:2021yyh,Butter:2021csz}. The implementation follows 
the variational approximation substituting the linear layer with
a mixture of uncorrelated Gaussians with learnable 
means and a diagonal covariance matrix. In practice, we only 
upgrade the last layer of each sub-network to a Bayesian layer~\cite{sharma2023bayesian}. 

\begin{figure}[b!]
    \centering
    \includegraphics[width=0.49\textwidth]{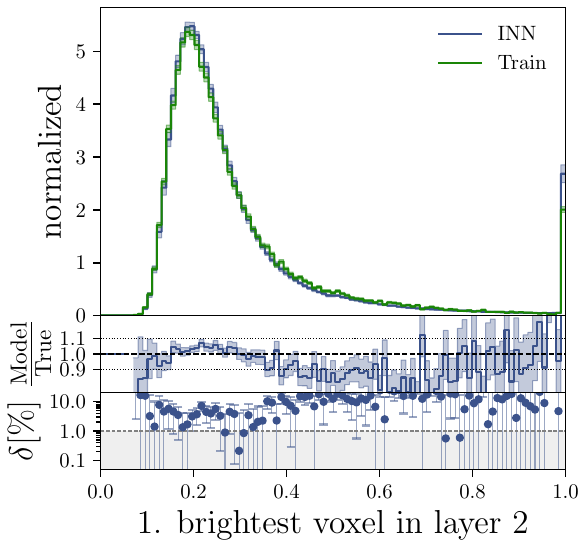}
    \includegraphics[width=0.49\textwidth]{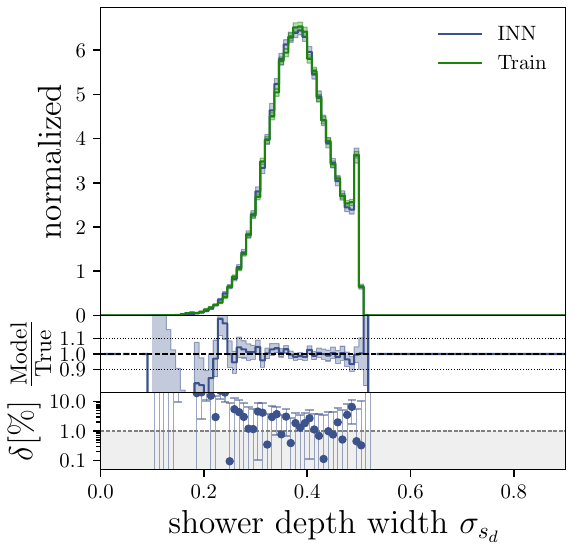}
    \caption{Comparison between CaloINN and \geant\ on two high level features.
            Brightest voxel distribution in layer-1 (left), and width of the shower depth (bottom).
            Error bars on the INN are calculated after sampling from the Bayesian network
            $N=50$ times.}
    \label{fig:bay_hlf}
\end{figure}

Figure~\ref{fig:bay_hlf} showcases two high--level features as examples of the performance of the CaloINN as compared to the training data  distribution generated by \geant. We show 
the brightest voxel distribution in layer 0 and the width of the shower 
depth width defined as the standard deviation of $s_d$~\cite{Krause:2021ilc}, with 
\begin{equation}
    s_d = \frac{\sum_{k=0}^2 k E_k}{\sum_{k=0}^2 E_k}.
\end{equation}
The error bars in the \geant\ distribution are 
the statistical errors while for the INN we estimate the 
uncertainties by sampling $N=50$ times from the network and 
resampling the network parameters each time.

\begin{figure}
    \centering
    \includegraphics[width=0.45\linewidth,]{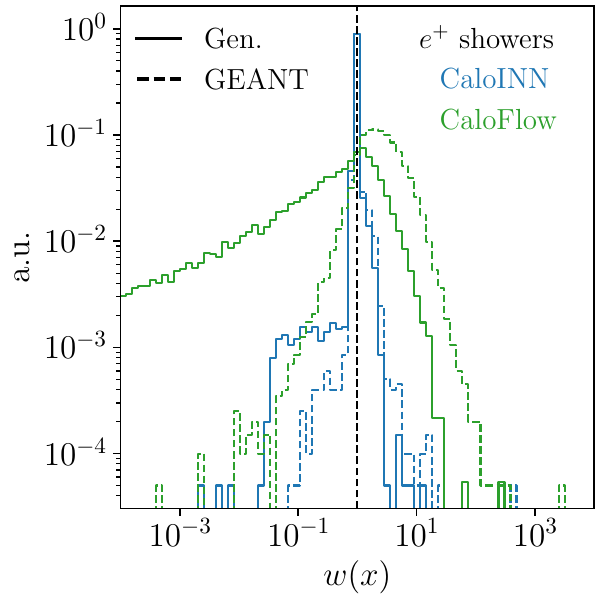}
    \includegraphics[width=0.45\linewidth]{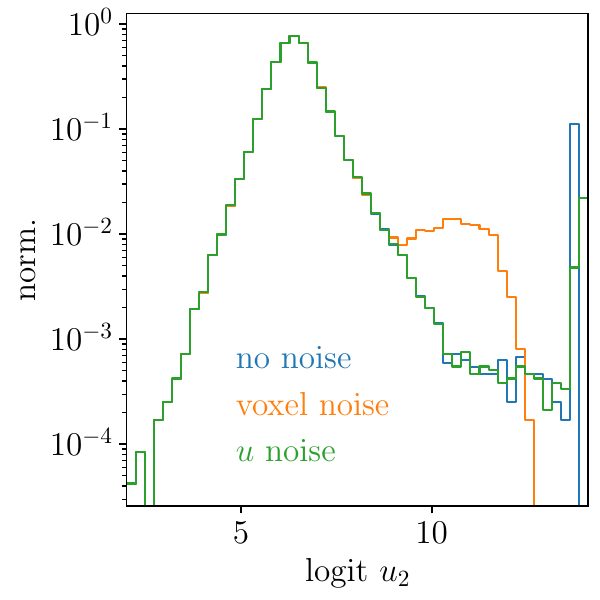}
    \caption{(left) Weight distribution of CaloFlow and CaloINN for $e^{+}$ showers. (right) $u_2$ distribution with different noise injections.}
    \label{fig:cls_gan}
\end{figure}

To evaluate our networks on low-level observables, we resort again
to classifier-based metrics. As already studied in a previous work \cite{Das:2023ktd},
the INN samples are indistinguishable from the \geant\ counterpart besides a few 
specific phase-space regions.
We train a classifier on the CaloFlow samples and find a large tail towards small weights.
From clustering of the tail, we observe a clear dependence on the energy deposition
total energy deposition. We link this effect to the learned energy variable $u_2=E_1/(E_1+E_2)$
and the noise injection procedure. If the noise is added at voxel-level, before calculating
the additional energy variables, the flow learns distorted energy ratio distributions.
Especially in the last layer, where the average energy deposition is smaller, this effect is larger.
We summarize this effect in Fig.~\ref{fig:cls_gan}. We also provide the AUCs and the generation timings in Tab.~\ref{tab:aucs_gen}.

\begin{table}[ht!]
\centering
\begin{tabular}{ll@{\hskip 10pt}l@{\hskip 10pt}l} \toprule
\multicolumn{2}{l}{AUC}            & CaloFlow\cite{Krause:2021ilc}      & CaloINN \\ \midrule
\multirow{3}{*}{$e^{+}$}  & unnorm.  &  0.859(10)    &   0.525(2)      \\
                        & norm.    &  0.870(2)     &   0.598(3)      \\
                        & hlf      &  0.795(1)     &   0.656(2)      \\ \midrule
\multirow{3}{*}{$\gamma$}  & unnorm.  &  0.756(50)    &   0.530(2)      \\
                        & norm.    &  0.796(2)     &   0.584(2)      \\
                        & hlf      &  0.727(2)     &   0.671(2)      \\ \midrule
\multirow{3}{*}{$\pi^{+}$} & unnorm.  &  0.649(3)     &   0.662(2)      \\
                        & norm.    &  0.755(3)     &   0.735(4)      \\
                        & hlf      &  0.888(1)     &   0.786(4)      \\ \bottomrule
\end{tabular}
\begin{tabular}{l|l@{\hskip 10pt}l@{\hskip 10pt}l} \toprule
   & Batch size & CaloFlow~\cite{Krause:2021wez} & CaloINN \\ \midrule
\multirow{3}{*}{GPU} & 1       & $ 55.12\pm 0.19 {}^*$ & $ 23.79\pm 0.10 {}^*$  \\
                     & 100     & $ 0.744\pm 0.04$ & $ 0.425\pm 0.005$    \\
                     & 10000   & $ 0.249\pm 0.003$ & $ 0.211\pm 0.003$   \\ \midrule
\multirow{3}{*}{CPU} & 1       & $119.9 \pm 0.9 {}^*$ & $46.39 \pm 3.18 {}^*$  \\
                     & 100     & $ 3.13\pm 0.11$ & $ 1.14\pm 0.03$  \\
                     & 10000   & $ 1.681\pm 0.004$ & $ 0.72\pm 0.01$   \\  \bottomrule
\end{tabular}
\caption{(left) AUC of the two classifiers trained on the CaloFlow teacher and CaloINN samples. (right) Per shower generation timings in ms. We show mean and standard deviation of 10 independent runs of generating 100k showers. The star indicates that only 10k samples were generated in total.}
\label{tab:aucs_gen}
\end{table}

\clearpage
\subsection{Additional histograms}
\label{app:inc_hists}

\begin{figure}[b!]
    \includegraphics[width=0.33\textwidth]{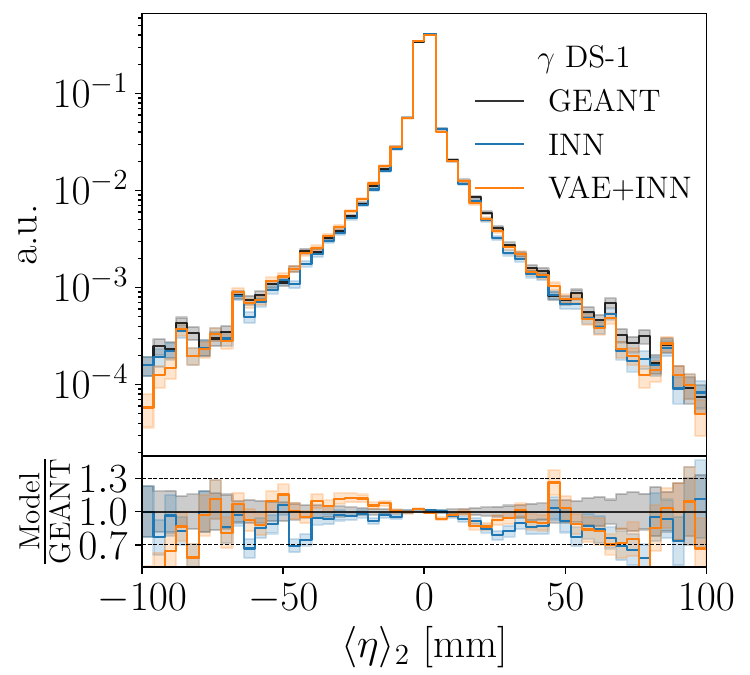}
    \includegraphics[width=0.33\textwidth]{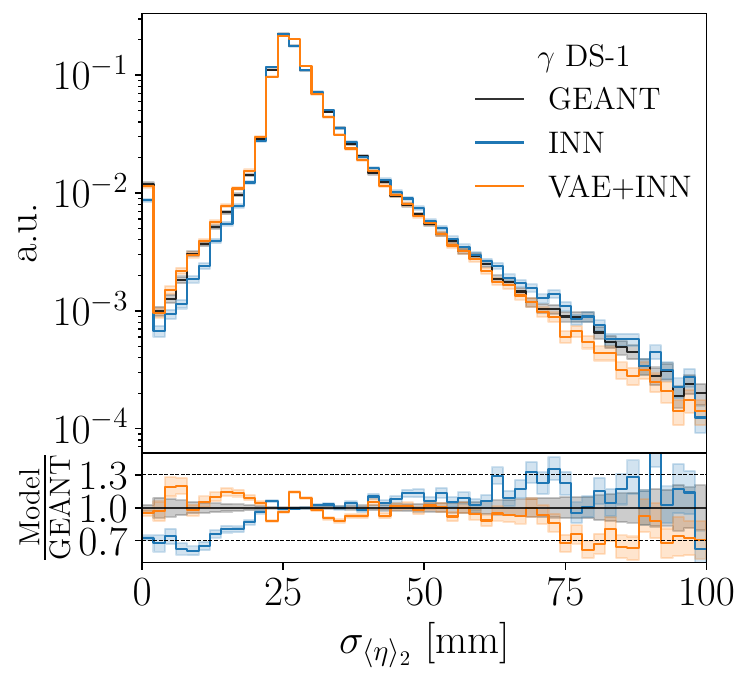}
    \includegraphics[width=0.33\textwidth]{figs/ds1-photons/Sparsity_layer_2_dataset_1-photons} \\
    \includegraphics[width=0.33\textwidth]{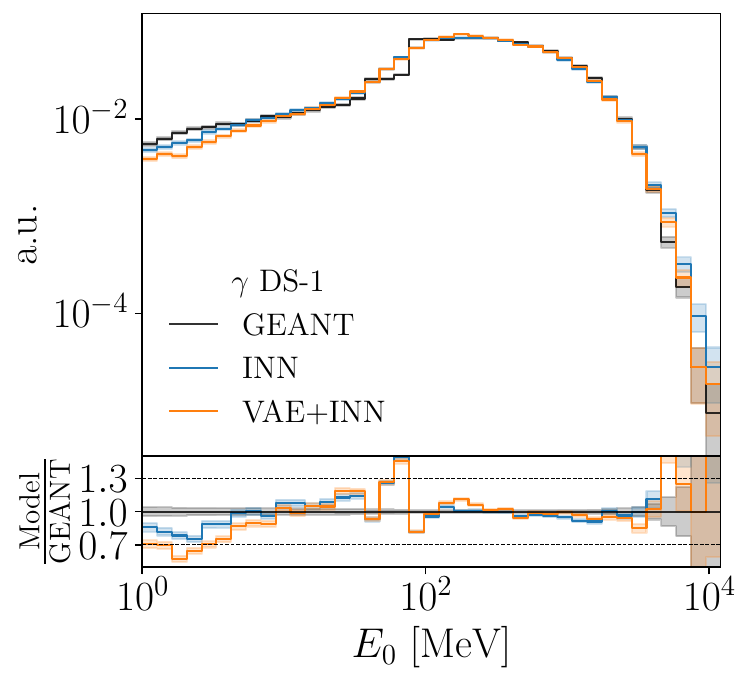}
    \includegraphics[width=0.33\textwidth]{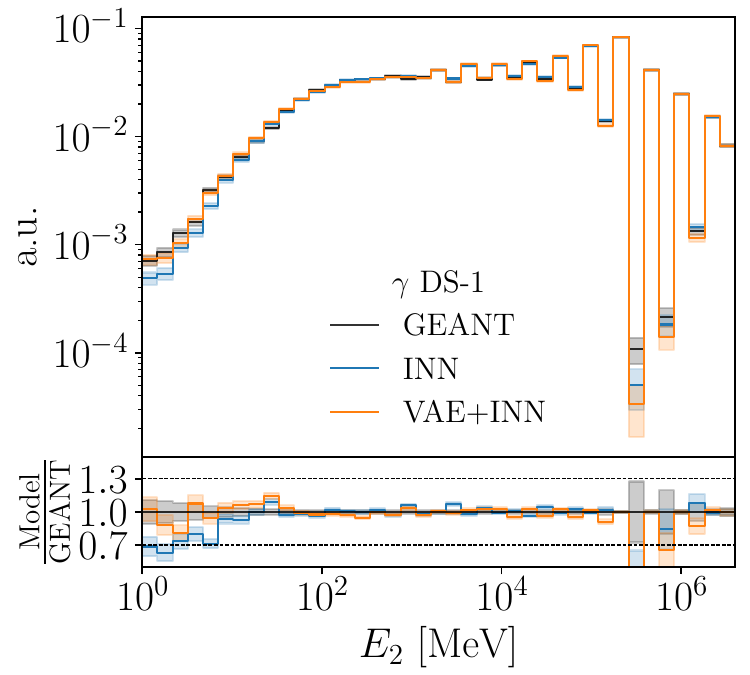}
    \includegraphics[width=0.33\textwidth]{figs/ds1-photons/Etot_Einc_dataset_1-photons}
    \caption{Set of high-level features for $\gamma$ showers in
      dataset~1 inclusive in $E_\text{inc}$, compared between \geant, INN, and VAE+INN.}
\end{figure}

\begin{figure}[b!]
    \includegraphics[width=0.33\textwidth]{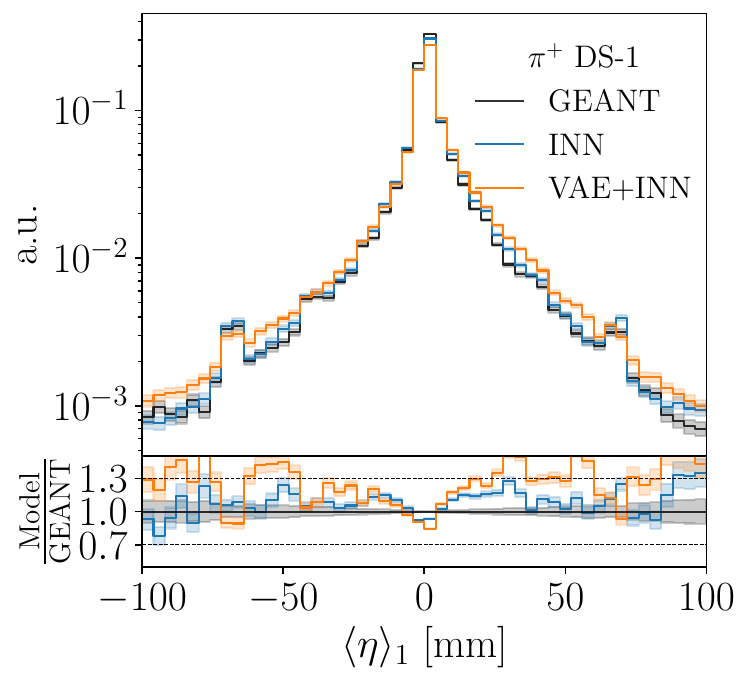}
    \includegraphics[width=0.33\textwidth]{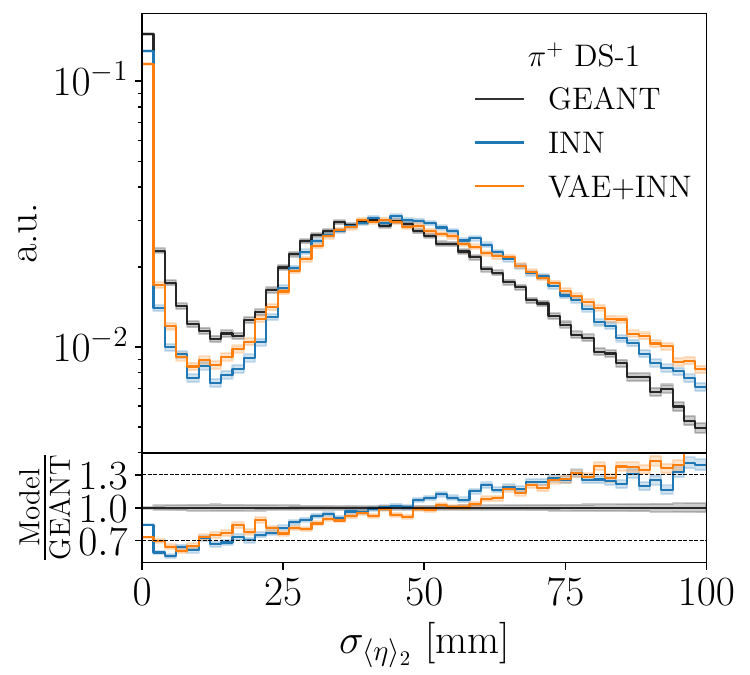}
    \includegraphics[width=0.33\textwidth]{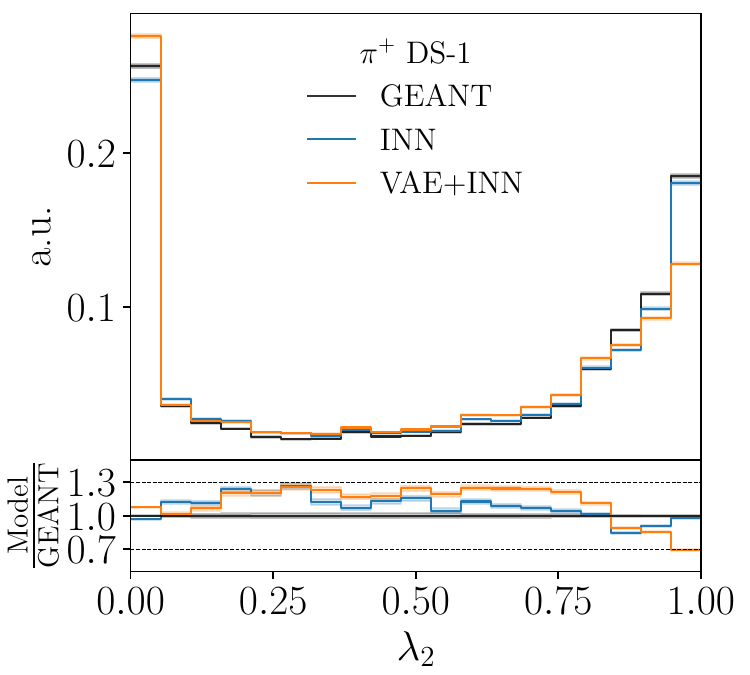} \\
    \includegraphics[width=0.33\textwidth]{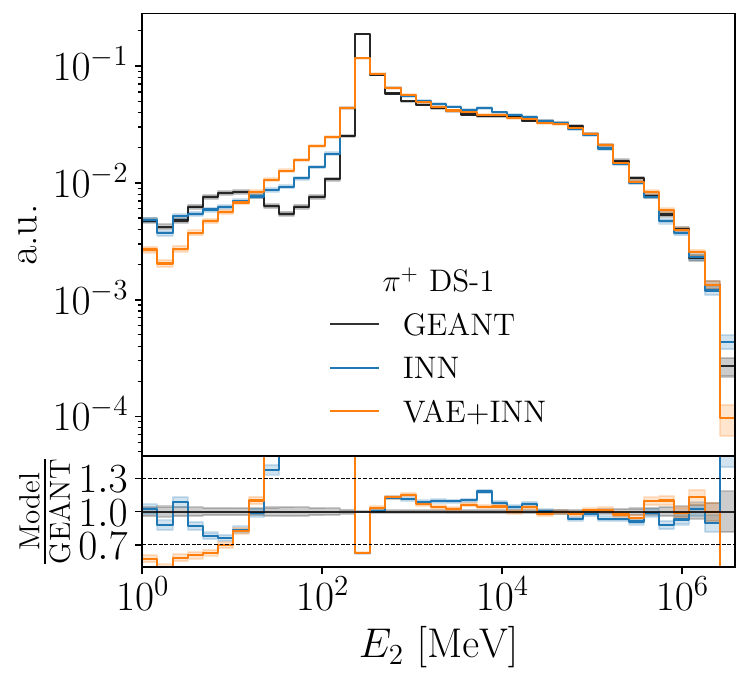}
    \includegraphics[width=0.33\textwidth]{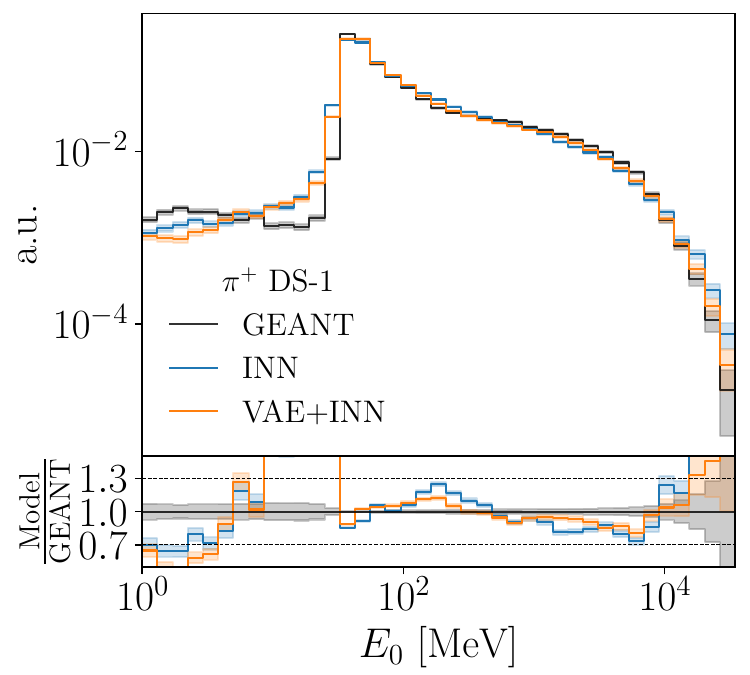}
    \includegraphics[width=0.33\textwidth]{figs/ds1-pions/Etot_Einc_dataset_1-pions}
    \caption{Set of high-level features for pion showers in dataset~1 inclusive in $E_\text{inc}$,
      compared between \geant, INN, and VAE+INN.}
\end{figure}

\begin{figure}[t!]
    \centering
    \includegraphics[width=\textwidth]{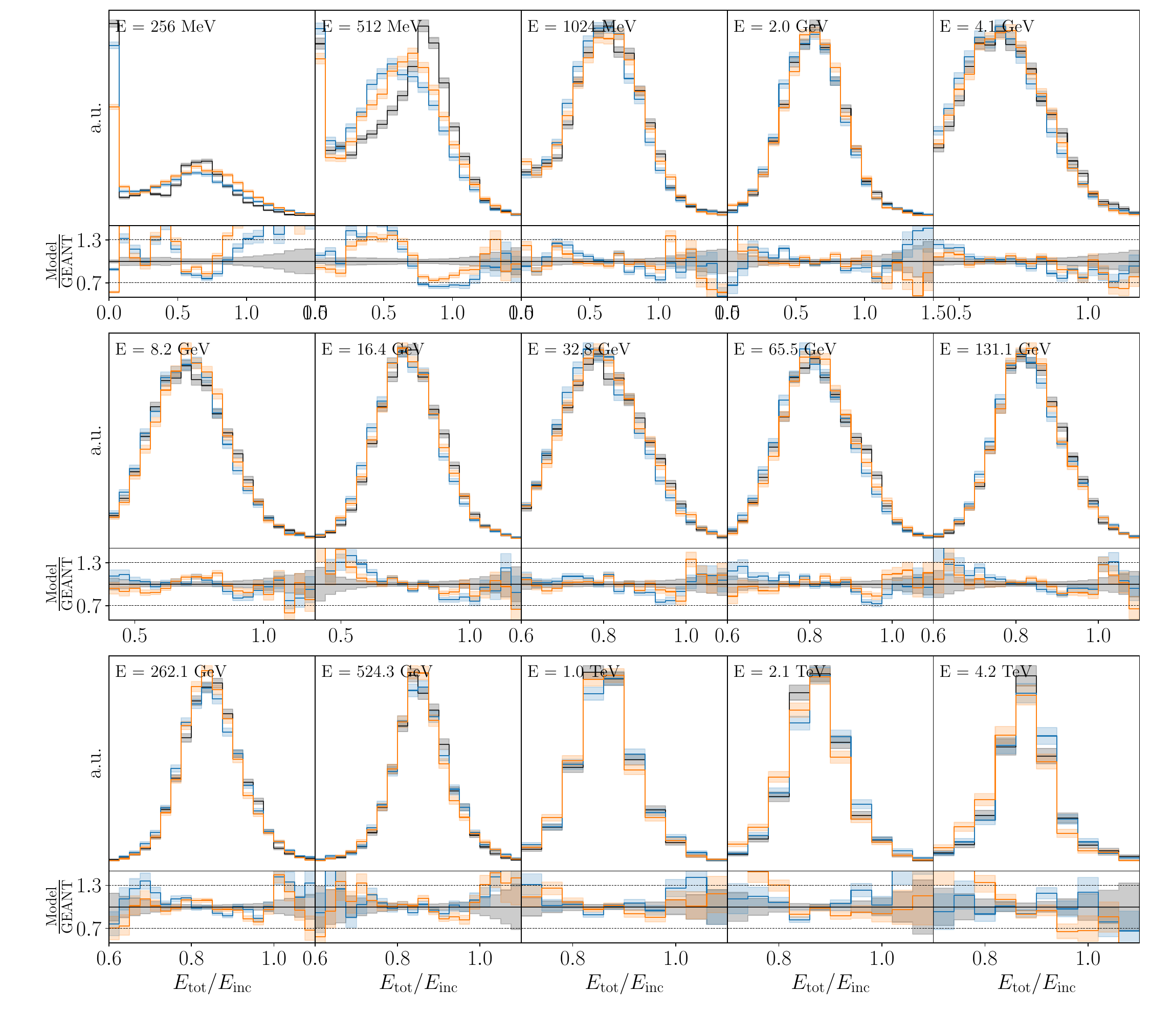}
    \caption{Energy ratio $E_\text{tot}/E_\text{inc}$ for each
      discrete incident energy, compared between \geant, INN, and
      VAE+INN for $\pi^{+}$ showers.  }
\end{figure}

\begin{figure}[t]
    \includegraphics[width=0.33\textwidth, page=21]{figs/ds2/ECEta_layer_dataset_2.pdf}
    \includegraphics[width=0.33\textwidth, page=21]{figs/ds2/WidthEta_layer_dataset_2.pdf}
    \includegraphics[width=0.33\textwidth, page=21]{figs/ds2/Sparsity_layer_dataset_2.pdf} \\
    \includegraphics[width=0.33\textwidth, page=1]{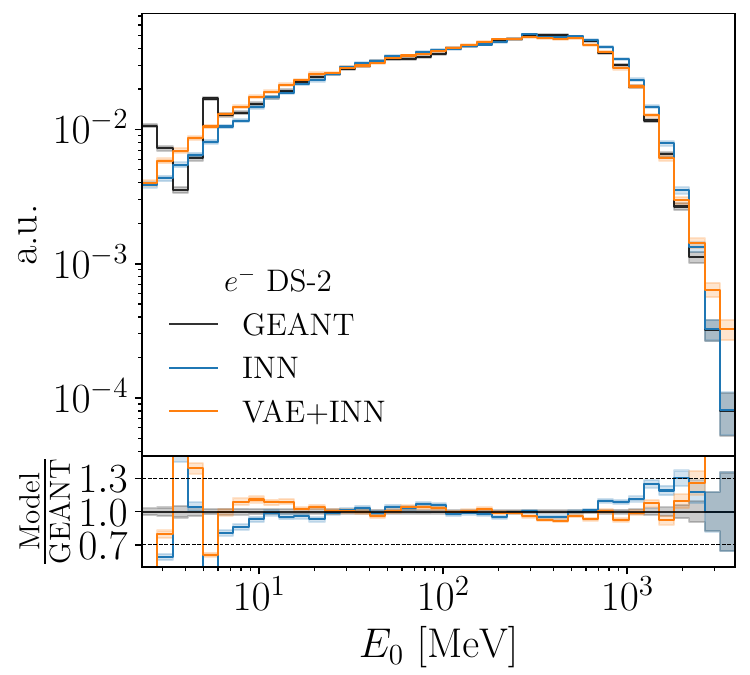}
    \includegraphics[width=0.33\textwidth, page=21]{figs/ds2/E_layer_dataset_2.pdf}
    \includegraphics[width=0.33\textwidth]{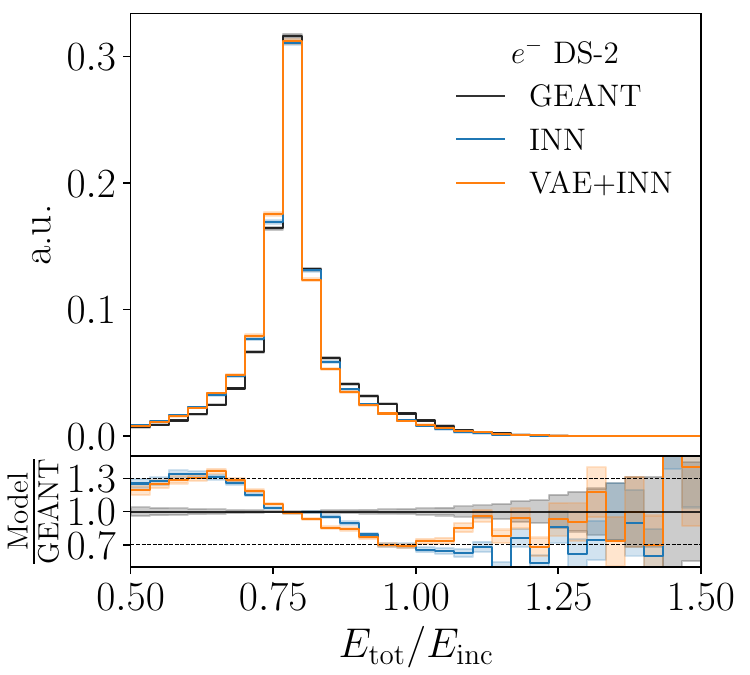}
    \caption{Set of high-level features for electron showers in
      dataset~2 inclusive in $E_\text{inc}$, compared between \geant, INN, and VAE+INN.}
\end{figure}

\clearpage
\bibliographystyle{SciPost-bibstyle-arxiv}
\bibliography{literature}

\end{document}

%% file: incl_settings.tex
\usepackage[utf8]{inputenc} 
\usepackage[T1]{fontenc} 	
\usepackage[english]{babel} 

\usepackage[bitstream-charter]{mathdesign}
\usepackage{geometry} 		
\usepackage{amsmath} 		
\usepackage{amsthm} 		
\usepackage{mathtools} 		
\usepackage{float} 			
\usepackage{graphicx} 		
\usepackage{tabularx} 		
\usepackage{booktabs} 		
\usepackage{color, xcolor} 	
\usepackage{pdfpages} 		
\usepackage{extarrows} 		
\usepackage{multirow} 		
\usepackage{multicol} 		
\usepackage{caption} 		
\usepackage{subcaption} 	
\usepackage{enumitem} 		
\usepackage{setspace} 		
\usepackage{xspace} 		
\usepackage{ragged2e} 		
\usepackage{stackrel} 		
\usepackage{tikz} 			
\usepackage{braket} 		
\usepackage{bm} 			
\usepackage{tensor} 		
\usepackage{slashed} 		
\usepackage{siunitx} 		
\usepackage{lastpage} 		
\usepackage{cite} 			
\usepackage[normalem]{ulem} 
\usepackage{fontawesome} 	
\usepackage{tocloft} 		
\usepackage{titlesec} 		
\usepackage{doi} 			
\usepackage{hyperref} 		
\usepackage[most]{tcolorbox} 					
\usepackage[nameinlink, capitalize]{cleveref} 	
\usepackage[nottoc, notlot, notlof]{tocbibind} 	
\usepackage[ruled, vlined]{algorithm2e} 		

\usepackage{array, multirow, bigdelim, makecell, booktabs}

\hypersetup{
	pdftitle={Normalizing Flows for High-Dimensional Detector Simulations},
	pdfauthor={Florian Ernst, Luigi Favaro, Claudius Krause, Tilman Plehn, and David Shih},
	colorlinks=true, 			
	linkcolor={red!50!black}, 	
	citecolor={blue!50!black}, 	
	urlcolor={blue!80!black} 	
} 

\DeclareSymbolFont{usualmathcal}{OMS}{cmsy}{m}{n}
\DeclareSymbolFontAlphabet{\mathcal}{usualmathcal}

\SetArgSty{textnormal}
\SetKwComment{Comment}{{\small\#}~}{}
\SetCommentSty{mycommfont}

\setitemize{itemsep=0pt, parsep=0pt} 				
\setenumerate{itemsep=0pt, parsep=0pt} 				
\setitemize{itemsep=2pt,topsep=2pt,parsep=0pt,partopsep=0pt,leftmargin=*}
\setenumerate{itemsep=0pt,topsep=2pt,parsep=0pt,partopsep=0pt,labelindent=3pt,leftmargin=*}
\usepackage{scalerel,stackengine,amsmath}

\usetikzlibrary{shapes.geometric, arrows, positioning, calc}

\tikzstyle{crc} = [circle, rounded corners=0.3ex, minimum width=1.5cm, minimum height=1cm, text centered, align=center, inner sep=0, fill=white, font=\LARGE, draw]

%% file: incl_shortcuts.tex
\newcommand{\kl}{D_\text{KL}}

\newcommand{\pl}{p_\text{latent}}
\newcommand{\pd}{p_\text{data}}

\newcommand{\pmd}{p_\text{model}}

\newcommand{\loss}{\mathcal{L}}

\newcommand{\XLangle}{\Big\langle}
\newcommand{\XRangle}{\Big\rangle}
\newcommand{\XXLangle}{\Bigg\langle}
\newcommand{\XXRangle}{\Bigg\rangle}

\newcommand\one{\leavevmode\hbox{\small1\normalsize\kern-.33em1}}

\newcommand{\qqquad}{\qquad \qquad}

\def\slashchar#1{\setbox0=\hbox{$#1$}           
   \dimen0=\wd0                                 
   \setbox1=\hbox{/} \dimen1=\wd1               
   \ifdim\dimen0>\dimen1                        
      \rlap{\hbox to \dimen0{\hfil/\hfil}}      
      #1                                        
   \else                                        
      \rlap{\hbox to \dimen1{\hfil$#1$\hfil}}   
      /                                         
   \fi}

\definecolor{orange}{RGB}{255,163,0}

\newcommand\geant{\textsc{Geant}4\xspace}

\definecolor{lightblue}{RGB}{208,232,242}
\definecolor{lightgreen}{RGB}{213,232,212}
\definecolor{lightyellow}{RGB}{255,242,204}
\definecolor{lightlavender}{RGB}{225,213,231}
\definecolor{lightcoral}{RGB}{248,206,204}
\definecolor{lightgrey}{RGB}{230,230,230}
\definecolor{lightpeach}{RGB}{255,230,204}
\definecolor{lightred}{RGB}{255,200,200}

%% file: figs/networks/inn.tex
\begin{tikzpicture}[node distance=1cm, line width=2pt, font=\LARGE]

\coordinate (A);
\coordinate (B) at ([xshift=+7cm] A);
\coordinate (C) at ([yshift=+10mm] B);
\coordinate (D) at ([yshift=+10mm] A);

\draw[fill=lightgrey] (A) -- (B) -- (C) -- (D) -- cycle;

\node at ($(A)!0.5!(B) + (0, 0.4cm)$) {$x$, $u$};

\coordinate (A'') at ([xshift = 2.cm, yshift=-2cm] A);
\coordinate (B'') at ([xshift=+3.0cm] A'');
\coordinate (C'') at ([yshift=-2cm] B'');
\coordinate (D'') at ([yshift=-2cm] A'');

\draw[fill=lightgrey] (A'') -- (B'') -- (C'') -- (D'') -- cycle;

\coordinate (A') at ([yshift=-13cm] A);
\coordinate (B') at ([yshift=-13cm] B);
\coordinate (C') at ([yshift=-13cm] C);
\coordinate (D') at ([yshift=-13cm] D);

\draw[fill=lightgrey] (A') -- (B') -- (C') -- (D') -- cycle;

\node at ($(A')!0.5!(B') + (0, 0.4cm)$) {$\hat{x}$, $\hat{u}$};

\draw[fill=lightblue] ([yshift=-15mm]A) -- ([yshift=-15mm]B) -- ([yshift=-15mm]C) -- ([yshift=-15mm]D) -- cycle;

\draw[fill=lightcoral] ([yshift=15mm]A') -- ([yshift=15mm]B') -- ([yshift=15mm]C') -- ([yshift=15mm]D') -- cycle;

\draw[fill=lightblue] ([yshift=30mm]A') -- ([yshift=30mm]B') -- ([yshift=30mm]C') -- ([yshift=30mm]D') -- cycle;

\draw[->] ([xshift =5mm, yshift=-18mm]A) -- ([xshift=5mm, yshift=-8.8cm]A);
\draw[->] ([xshift =-4mm, yshift=-18mm]B) -- ([xshift=-4mm, yshift=-8.8cm]B);

\node (trafo) [crc, below of = C, yshift=-7.3cm, xshift=-4mm, font=\Large, fill=lightgreen]{$f$};

\draw[->] ([xshift =5mm, yshift=-5cm]A) -- (trafo);

\draw[fill=lightcoral] ([yshift=-1.2cm]D'') -- ([yshift=-1.2cm]C'') -- ([yshift=-3.3cm]C'') -- ([yshift=-3.3cm]D'') -- cycle;

\draw[->] ([xshift =1.5cm, yshift=-0.1cm]D'') -- ([xshift=1.5cm, yshift=-1.1cm]D'');

\node at ($(A'')!0.5!(B'') - (0, 1.0cm)$) {$\log E_{\text{inc}}$};

\node[text width=1.6cm] at ($(A'')!0.5!(B'') - (0, 4.3cm)$) {Spline params.};

\node at ($(A)!0.5!(B) - (0, 1.0cm)$) {Split};

\node at ($(A')!0.5!(B') + (0, 2.0cm)$) {ActNorm};

\node at ($(A')!0.5!(B') + (0, 3.5cm)$) {Permute};

\end{tikzpicture}

%% file: figs/networks/vae-inn.tex
\begin{tikzpicture}[node distance=1cm, line width=2pt, font=\LARGE]

\coordinate (A);
\coordinate (B) at ([xshift=7cm] A);
\coordinate (C) at ($([yshift=-4cm]A) !0.5! ([yshift=-4cm]B) + (1,0)$);
\coordinate (D) at ($([yshift=-4cm]A) !0.5! ([yshift=-4cm]B) + (-1,0)$);

\draw[fill=lightcoral] (A) -- (B) -- (C) -- (D) -- cycle;

\coordinate (A') at ([yshift=-11cm]A);
\coordinate (B') at ([yshift=-11cm]B);
\coordinate (C') at ($([yshift=4cm]A') !0.5! ([yshift=4cm]B') + (1,0)$);
\coordinate (D') at ($([yshift=4cm]A') !0.5! ([yshift=4cm]B') + (-1,0)$);

\draw[fill=lightcoral] (A') -- (B') -- (C') -- (D') -- cycle;

\draw[fill=lightblue] ([yshift=-5mm]D) -- ([yshift=-5mm]C) -- ([yshift=5mm]C') -- ([yshift=5mm]D') -- cycle;

\draw[fill=lightgreen] 
([yshift=-5mm, xshift=5cm]D) -- 
([yshift=-5mm, xshift=5cm]C) -- 
([yshift= 5mm, xshift=5cm]C') -- 
([yshift= 5mm, xshift=5cm]D') -- 
cycle;

\draw[->]
([yshift=-10mm, xshift=5mm]C) -- 
([yshift=-10mm, xshift=4.5cm]D) node[midway,below, yshift=-1mm] {\textbf{INN}};

\draw[->]
([yshift=10mm, xshift=4.5cm]D') --
([yshift=10mm, xshift=5mm]C');

\node[align=center] at ($(A)!0.5!(B) + (0,-1.3cm)$) {\textbf{Encoder}\\
$E(z|x)$};
\node[align=center] at ($(A')!0.5!(B') + (0,1.3cm)$) {\textbf{Decoder}\\
$D(x|z)$};

\node at ($(D)!0.5!(C) - (0,1.5cm)$) {$Z, u$};
\node at ([xshift=5cm]$(D)!0.5!(C) - (0,1.4cm)$) {$\tilde Z$};

\draw[fill=lightgrey] 
([yshift=5mm]A) -- 
([yshift=5mm]B) -- 
([yshift=15mm]B) -- 
([yshift=15mm]A) -- 
cycle;

\draw[fill=lightgrey] 
([yshift=-5mm]A') -- 
([yshift=-5mm]B') -- 
([yshift=-15mm]B') -- 
([yshift=-15mm]A') -- 
cycle;

\draw[fill=lightgrey] ([yshift=-5mm, xshift=-35mm]D) -- ([yshift=-5mm, xshift=-35mm]C) -- ([yshift=5mm, xshift=-35mm]C') -- ([yshift=5mm, xshift=-35mm]D') -- cycle;

\draw[->] ([xshift=-25mm]D) -- ([yshift=15mm, xshift=-25mm]D) |- ([yshift=15mm, xshift=-15mm]D);

\draw[->] ([xshift=-25mm]D') -- ([yshift=-15mm, xshift=-25mm]D') |- ([yshift=-15mm, xshift=-15mm]D');

\draw[->] ([xshift = -10mm, yshift=-15mm]D) -- ([xshift=-5mm, yshift=-15mm]D);

\node at ([xshift=-35mm]$(D)!0.5!(C) - (0,1.5cm)$) {$u$};

\node at ($(A)!0.5!(B) + (0, 1cm)$) {$x$};
\node at ($(A')!0.5!(B') - (0, 1cm)$) {$\hat{x}$};

\end{tikzpicture}